\documentclass[12pt,a4paper]{article}
\pdfoutput=1
\usepackage{jcappub}
\usepackage{natbib}
\usepackage{graphicx}
\usepackage[english]{babel}
\usepackage{color}
\usepackage{multirow}
\usepackage{ulem}
\usepackage{url}

\def\bea{\begin{eqnarray}}
\def\eea{\end{eqnarray}}
\def\be{\begin{equation}}
\def\ee{\end{equation}}

\newcommand{\g}{$\gamma$}

\title{Dark matter in the Reticulum II dSph: a radio search}

\author[a,b]{Marco Regis,}
\emailAdd{regis@to.infn.it}
\affiliation[a]{Dipartimento di Fisica, Universit\`{a} di Torino, via P. Giuria 1, I--10125 Torino, Italy}
\affiliation[b]{Istituto Nazionale di Fisica Nucleare, Sezione di Torino, via P. Giuria 1, I--10125 Torino, Italy}

\author[c,d]{Laura Richter}
\emailAdd{llrichter@gmail.com}
\affiliation[c]{SKA South Africa, 3rd Floor, The Park, Park Road, Pinelands, 7405, South Africa}
\affiliation[d]{Physics  and Electronics Department, Rhodes University, Grahamstown, 6140, South Africa}

\author[e]{and Sergio Colafrancesco}
\emailAdd{sergio.colafrancesco@wits.ac.za}
\affiliation[e]{School of Physics, University of the Witwatersrand, Johannesburg, South Africa}

\abstract{
We present a deep radio search in the Reticulum II dwarf spheroidal (dSph) galaxy performed with the Australia Telescope Compact Array.
Observations were conducted at 16 cm wavelength, with an rms sensitivity of 0.01 mJy/beam, and with the goal of searching for synchrotron emission induced by annihilation or decay of weakly interacting massive particles (WIMPs).
Data were complemented with observations on large angular scales taken with the KAT-7 telescope.
We find no evidence for a diffuse emission from the dSph and we derive competitive bounds on the WIMP properties. 
In addition, we detect more than 200 new background radio sources. Among them, we show there are two compelling candidates for being the radio counterpart of the possible \g-ray emission reported by other groups using Fermi-LAT data.
}

\date{\today}
\begin{document}
\maketitle

\section{Introduction}
\label{sec:Intro}

Dwarf spheroidal (dSph) galaxies are extremely important systems for cosmology and astrophysics.
They have been shown to be a promising avenue to test the hypothesis of weakly interacting massive particles (WIMPs) as dark matter (DM) candidates~\cite{Colafrancesco:2006he}.
dSphs are the faintest and most metal-poor stellar systems known. They are also the closest galaxies (other than the Milky-Way) and the most DM dominated objects in the local Universe (see, e.g., Ref.~\cite{McConnachie:2012vd} for a recent review).
Moreover, they lack recent star formation and show tight constraints on the presence of gas from HI observations~\cite{Grcevich:2009gt,Spekkens:2014}.

The dwarf galaxy Reticulum II (RetII) was discovered in 2015 in first year Dark Energy Survey data~\cite{Koposov:2015cua,Bechtol:2015}. 
Spectroscopic follow-­ups~\cite{Simon:2015fdw,Walker:2015,Koposov:2015} confirmed that the object is an ultra­-faint dSph galaxy, satellite of the Milky Way.
They found RetII to be strongly DM dominated with a measured velocity dispersion around 3.5 km/s and a mass­-to­-light ratio within its half­-light radius around 500 $M_\odot/L_\odot$. 
RetII is one of the most metal-­poor galaxies known with a mean metallicity of $[Fe/H]<­-2.5$. 
Moreover, recent chemical abundance determinations for the nine brightest red giant members of RetII surprisingly indicate high levels of $r$-process material~\cite{Ji:2016,Roederer:2016}, differently from every other ultra­-faint dSph, making RetII a unique target. 

We performed deep radio observations around the direction of RetII with a primary beam covering 23.7 arcmin. Data have been collected with the Australia Telescope Compact Array (ATCA) operating at 16~cm wavelength with an rms sensitivity around 10 $\mu$Jy.
We adopted a compact setup, optimal to test the presence of a diffuse radio continuum signal on the scale of a few arcminutes.
Nevertheless, the presence of a long-baseline allowed us to go deep in detecting background sources.

The observed properties of the stellar population in Reticulum II imply that the thermal and non-­thermal radio emissions from common astrophysical mechanisms are expected to be significantly below the detection threshold of current radio telescopes~\cite{Regis:2014koa}. 
On the other hand, if DM is in form of WIMPs and RetII hosts a non­-negligible magnetic field, a synchrotron radiation from the electrons injected by DM annihilations or decays can be within the reach.

Indeed, the proximity (RetII distance is $\sim 30$ kpc~\cite{Koposov:2015cua,Bechtol:2015}) and high-DM content imply a relevant expected WIMP emission.
RetII can have a large $J$-factor \footnote{The so-called $J$-factor is the angular and line-of-sight integral of the square of the DM density profile: $\int_{\Delta \Omega_{obs}}d\Omega\int_{l.o.s.}ds \,\rho_{DM}^2$.}, as shown in Refs.~\cite{Bonnivard:2015tta,Evans:2016xwx}.
The location in the sky is also ideal for WIMP searches, being well below the Galactic plane, in a region with low and uniform Galactic foreground.
The expected large DM signal, on one side, and the low astrophysical background, on the other, make RetII an ideal target for indirect searches of particle DM.
The half-­light radius of RetII is around 3.6 arcmin~\cite{Koposov:2015cua,Bechtol:2015} along its minor axis (with an ellipticity of 0.6). 
Our experimental setup aims at observing diffuse synchrotron radiation on scales ranging from one to fifteen arcmin, possibly associated to the DM halo. 

Various attempts have been pursued in the context of searching for WIMP-induced prompt emission of \g-rays or radiative emission (inverse Compton scattering in the X- and \g-ray bands and synchrotron radiation at radio frequencies) associated with WIMP-induced electrons and positrons in dSph (see, e.g., the Introduction in Ref.~\cite{Regis:2014tga} and references therein).
No evidence of diffuse signal has been robustly obtained so far at any relevant frequency and this only allowed to set upper limits on the DM annihilation/decay rate for a wide range of WIMP masses.

On the other hand, soon after RetII discovery, a possible \g-ray emission from the direction of the dSph was claimed in Ref.~\cite{Geringer-Sameth:2015lua}, where the authors show also that the derived signal is consistent with annihilation of DM particles (see also a possible evidence for a 511 keV emission in Ref.~\cite{Siegert:2016ijv}, and, on the other hand, multi-wavelength constraints in Ref.~\cite{Beck:2015rna}).
The statistical significance of the measurement is still unclear, and has been quoted at 0.43$\sigma$~\cite{Drlica-Wagner:2015xua}, 1.7$\sigma$~\cite{Fermi-LAT:2016uux}, 2$\sigma$~\cite{Zhao:2017pcz} and 3.7$\sigma$~\cite{Geringer-Sameth:2015lua}, depending on the analysis.
In this work, we study the possibility that the \g-ray emission is indeed real, but due to a blazar in the background of RetII, rather than to DM annihilations in the dSph.
To this aim, we extract radio sources from a region of interest (ROI) of 20 arcmin around RetII center, and cross-match these sources with infrared objects to identify possible blazars~\cite{Massaro:2012dh}.
Using an empirical radio-gamma relation~\cite{Ackermann:2015yfk}, we then predict the expected \g-ray emission in the RetII field.

Most recent and comprehensive searches for WIMPs with \g-ray observations of dSphs include Refs.~\cite{Ackermann:2015zua,Drlica-Wagner:2015xua,Ahnen:2016qkx,Li:2015kag,Fermi-LAT:2016uux}. They reach sensitivities at the level of the WIMP thermal relic annihilation cross section for DM masses lighter than 100 GeV.

Radio campaigns dedicated to DM searches in dSphs started few years ago, with two different approaches.
In Refs.~\cite{Spekkens:2013ik,Natarajan:2013dsa,Natarajan:2015hma}, observations were performed with a single-dish telescope (the Green Bank Telescope).
In Refs.~\cite{Regis:2014joa},\cite{Regis:2014koa} and \cite{Regis:2014tga} (hereafter R15a, R15b and R14, respectively), we made use of a radio interferometer (ATCA) that allows to resolve and subtract compact sources. The same approach is adopted in the present work.
The sensitivity of radio searches to WIMP signals depends on the assumptions on the magnetic properties of dSphs. Considering realistic scenarios, current telescopes can provide bounds comparable to \g-ray searches~\cite{Regis:2014tga}. The technique adopted here and in R14 can also be considered as preparatory for SKA and its precursor, where the sensitivity will dramatically increase, together with the ability to derive magnetic field properties through Faraday rotation and polarization measures.

The paper is organized as follows.
In Section~\ref{sec:obs}, the observing setup and data manipulation are summarized. 
We describe how we extracted sources and built the source catalog in Section~\ref{sec:cat}, where we also briefly discuss the main properties of the catalog.
In Section~\ref{sec:excess}, we investigate the possibility that one of the sources of the catalog is responsible for the \g-ray emission suggested by Ref.~\cite{Geringer-Sameth:2015lua}. 
In Section~\ref{sec:diffuse}, we search for an extended diffuse emission in RetII. Constraints on the WIMP DM parameter space are discussed in Section~\ref{sec:bound}.
Section~\ref{sec:concl} concludes.

\section{Observations and data reduction}
\label{sec:obs}

\begin{figure}[t]
   \centering
 \hspace{-0.8cm}
 \includegraphics[width=0.99\textwidth]{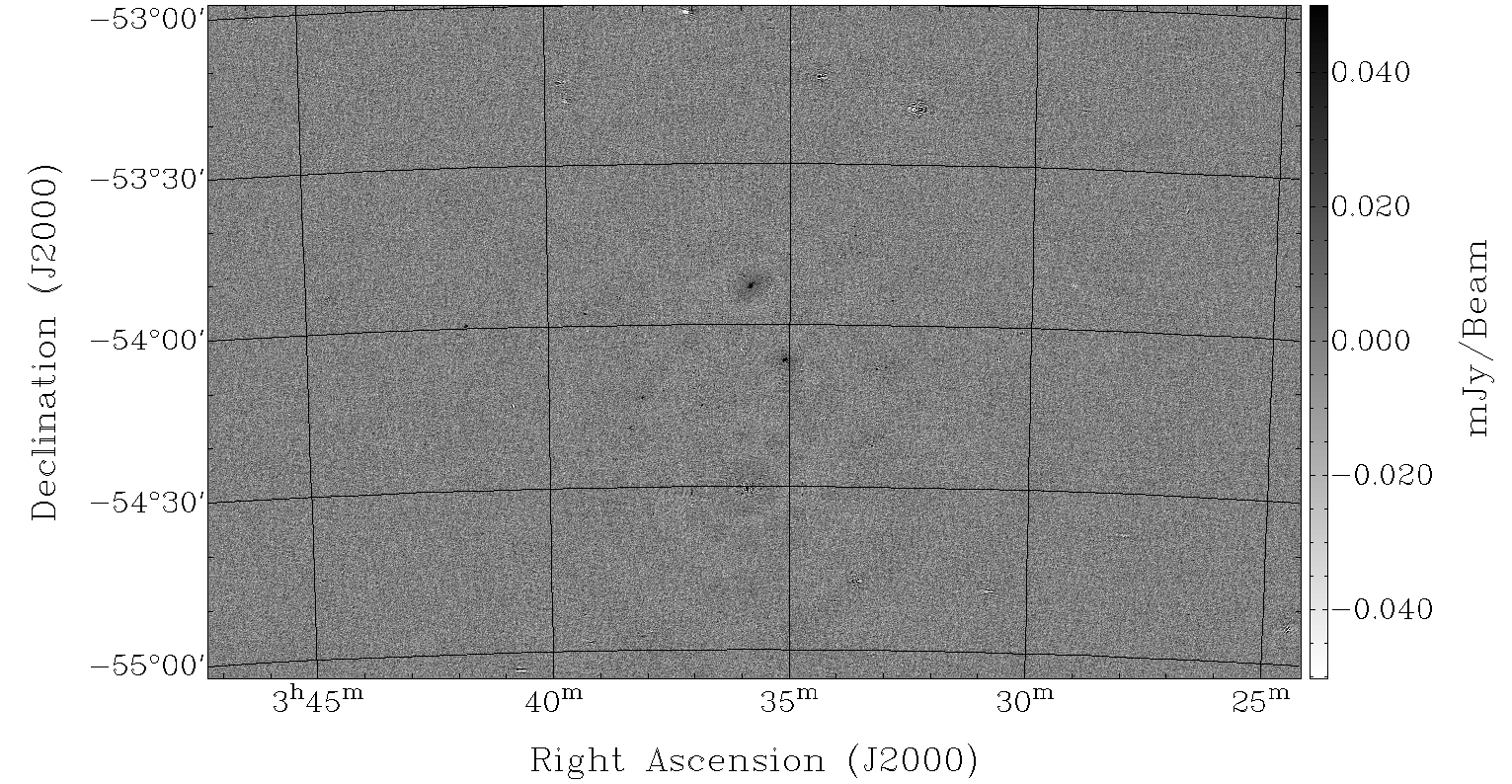}
\caption{Grayscale of the observational map obtained setting the robustness parameter to -1. All antennas are included and no tapering is applied. }
\label{fig:map_fulla}
\end{figure}

\begin{figure}[t]
\vspace{-2.5cm}
   \centering
 \hspace{-0.9cm}
   \includegraphics[width=0.99\textwidth]{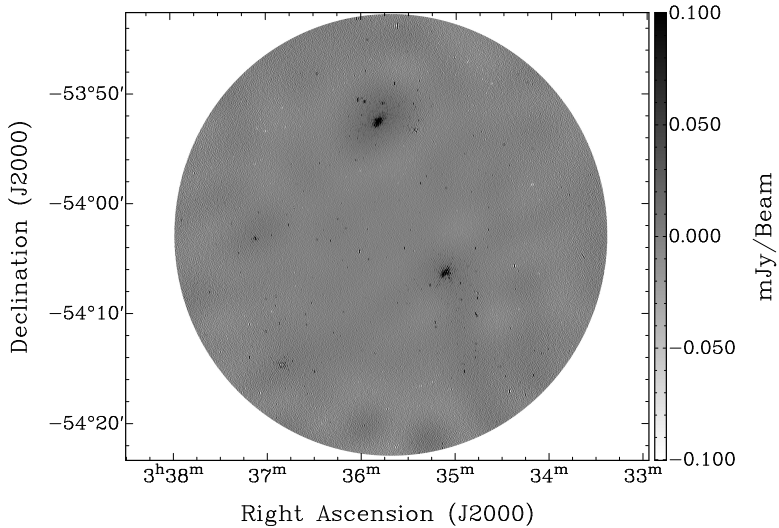}
\vspace{-6.5cm}
\caption{ Left: Grayscale of the region of interest considered in our analysis ($r_{-1}$ map). The map is corrected for primary beam effect. }
\label{fig:map_fullb}
\end{figure}

ATCA observations were performed during July 2016 with the six 22-m diameter antennae operating at 16 cm wavelength in the hybrid configuration H75. The total observing time amounts to 30 hours and the RetII pointing was centered at RA=$03$:$35$:$41$ and DEC=$-54$:$03$:$00$.
The observing setup is composed by both a compact array of five antennas (with a maximum baseline of 89 m) and long baselines involving a sixth antenna located at approximately 4.4 km from the core. 
At the center of the frequency range $1.1-3.1$ GHz, the primary beam corresponds to $23.7'$, while the synthesized beams are $7.5''\times 2.0''$ for the full array and $3.5'\times 3.0'$ if we do not include the long baselines involving the sixth antenna.

Short-spacings are required to detect extended emissions. With the compact array, we are sensitive to angular scales from 3 arcmin to a maximum size of well-imaged structures of about 15 arcmin (the length of the shortest baseline is 31 m that, at the center of the frequency band, corresponds to 18 arcmin).
With a three arcmin beam, however, the confusion limit is quickly approached.
The long baselines provide us with high-resolution mapping of the small-scale background sources. 

The data were reduced using the {\sc Miriad} data reduction package~\cite{Sault:95}. Calibration and imaging follow the procedures described in R15a.
We adopted the multi-frequency CLEAN algorithm~\citep{Sault:94}, a Briggs robustness parameter of -1 \citep{Briggs:thesis} and four iteration of self-calibration.
We produced three different maps.

\begin{figure}[t]
\vspace{-2.5cm}
   \centering
 \hspace{-0.7cm}
   \includegraphics[width=0.99\textwidth]{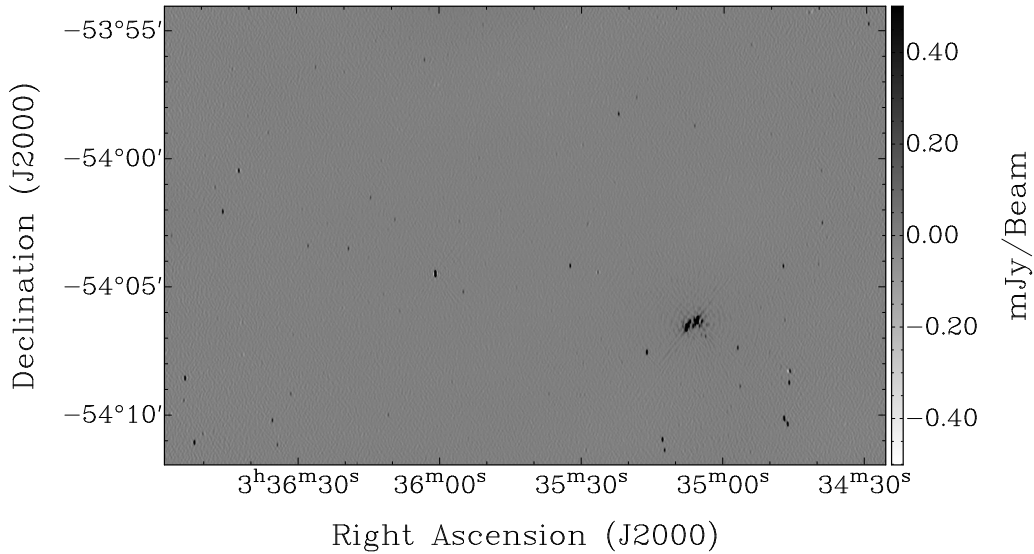}
\vspace{-6.5cm}
\caption{Zoom-in of the central region of the map in Fig.~\ref{fig:map_fullb}. }
\label{fig:map_zoom}
\end{figure}

\begin{figure}[t]
\vspace{-3cm}
\centering
 \begin{minipage}[htb]{8.8cm}
   \centering
 \includegraphics[width=\textwidth]{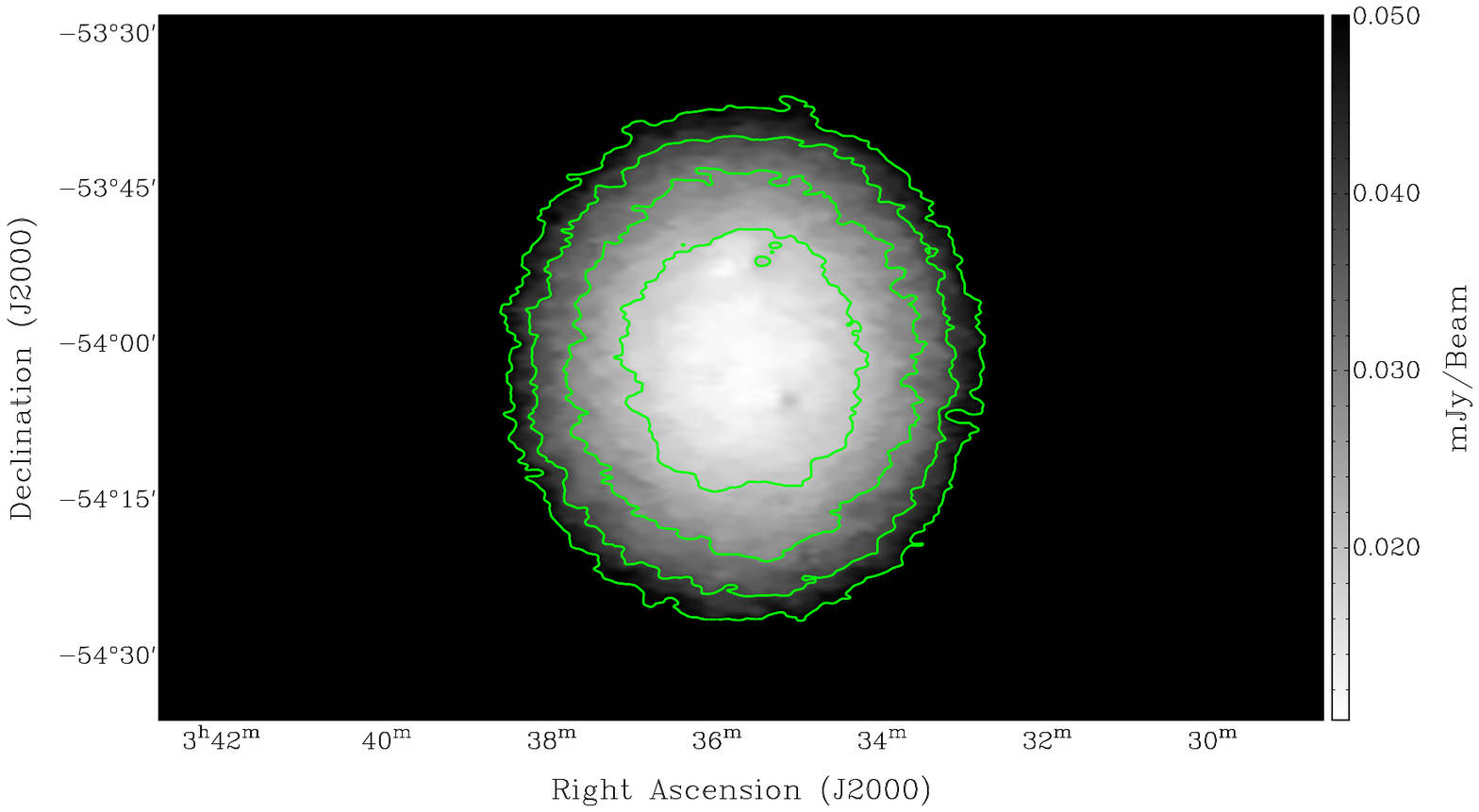}
 \end{minipage} 
 \begin{minipage}[htb]{6.5cm}
   \centering
\vspace{-3cm}
   \includegraphics[width=\textwidth]{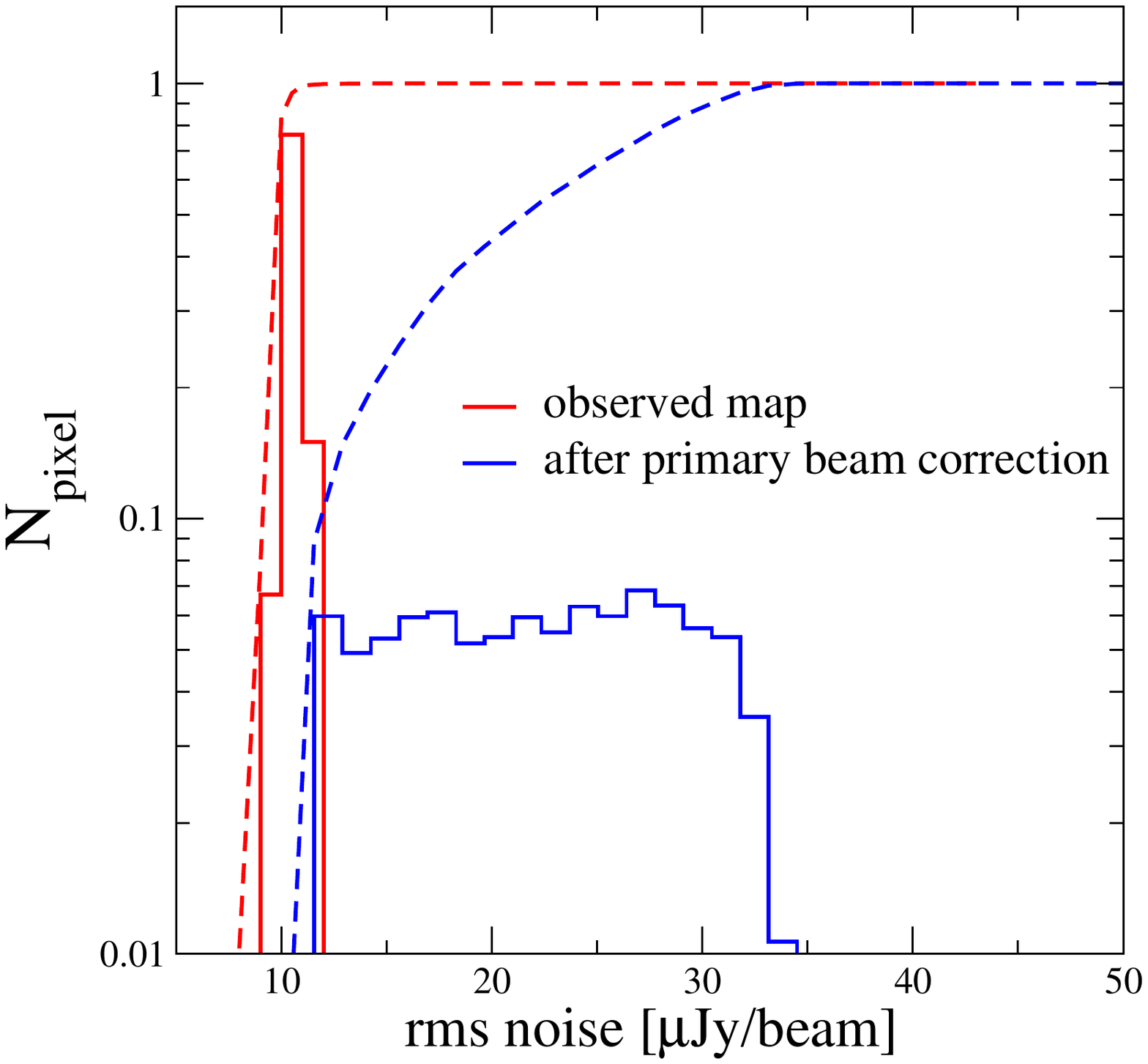}
 \end{minipage} 
\vspace{-3.5cm}
\caption{Left: RMS noise map derived with SExtractor. Contours start from 20 $\mu{\rm Jy}$ with steps of 10 $\mu{\rm Jy}$. The area within the primary beam has a nearly constant noise $<15\,\mu{\rm Jy}$. Right: Fraction of pixels in the map with a given noise (solid) and cumulative RMS distribution (dashed). Red (blue) curve refers to the map before (after) primary beam correction.}
\label{fig:noise}
\end{figure}

A high resolution map (we will refer to this as the $r_{-1}$ map) includes all the baselines and is shown in Fig.~\ref{fig:map_fulla} and \ref{fig:map_fullb}, with a zoom of the central region in Fig.~\ref{fig:map_zoom}.\footnote{All the maps presented in this paper can be retrieved at \url{http://personalpages.to.infn.it/~regis/c3103.html}.}
Its rms noise amounts to about 10 $\mu$Jy. In Fig.~\ref{fig:noise}, we show the structure of the rms noise as derived with the SEXTRACTOR package~\citep{Bertin:1996fj} (for more details, see R15a).
The $r_{-1}$ map shows a very good imaging of sources with size below a few arcsec. However, the reconstruction is not optimal for sources above approximately 10 arcsec.
This is due to the structure of the synthesized beam for the H75 antenna configuration. The gap in UV-coverage between the compact core and the UV-coverage from the 4 km baselines leads to a synthesized beam with large sidelobes, causing deconvolution errors for non-point sources.
We will come back to this point in the following Sections. 

We then produced a map excluding all the baselines involving the sixth antenna. It is named $no6$ map and is shown in Fig.~\ref{fig:map_kat}a. The resolutions significantly decreases but the map is sensitive to extended emissions.

\begin{figure}[t]
\vspace{-2.5cm}
   \centering
 \hspace{-0.9cm}
   \includegraphics[width=0.99\textwidth]{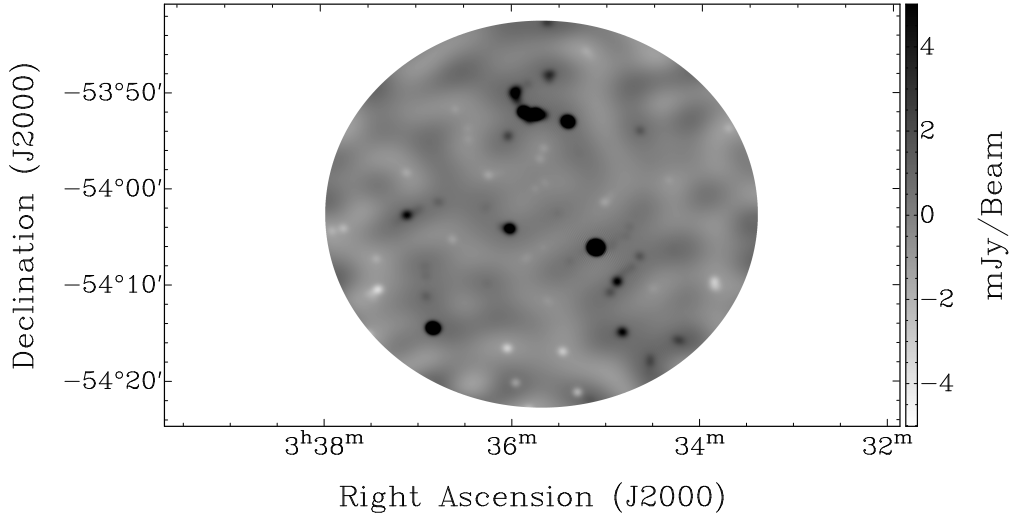}
\vspace{-6.5cm}
\caption{Map of the ROI tapered with FWHM=$15''$ ($f15$ map). The tapering is performed to downweight long baselines. }
\label{fig:map_tap}
\end{figure}

Finally, as a sort of compromise between the above two images, we apply a Gaussian taper of 15 arcsec, which effectively down-weights long baselines involving the sixth antenna, leading to a synthesized beam of $50''\times 47''$. The central region of this map, named $f15$, is shown in Fig.~\ref{fig:map_tap}.

\subsection{KAT-7 data}
RetII was observed also with the KAT-7 radio telescope~\cite{KAT7:2016} on 27 March 2016 and 7, 9, 11, 12 August 2016.
The March observation spanned 9 hours and 6 antennas were present. Only 5 antennas were present for the August time-slots, and the observations spanned 44 hours in total. RetII was observed in conjunction with bandpass calibrator 3C138 and gain calibrator 0302-623 (RA=$3$:$03$:$50.63$  DEC=$-62$:$11$:$25.5$).
The observations were centred on 1822 MHz, over 400 MHz split into 1024 channels, with 10 second dumps.

Each data set was calibrated in casapy using standard data reduction methods of bandpass calibration, gain calibration and absolute flux calibration. The calibrated data were jointly imaged using the casapy imager, with a robustness parameter of 0 and a noise threshold of 0.5 mJy. The confusion noise was higher than the thermal noise for the image, so the noise threshold was set by the confusion level evident in the image. The restoring beam was $3.6'\times 2.2'$,  position angle -27 degrees.

\begin{figure}[t]
   \centering
 \hspace{-0.8cm}
   \includegraphics[width=0.4\textwidth,angle=-90]{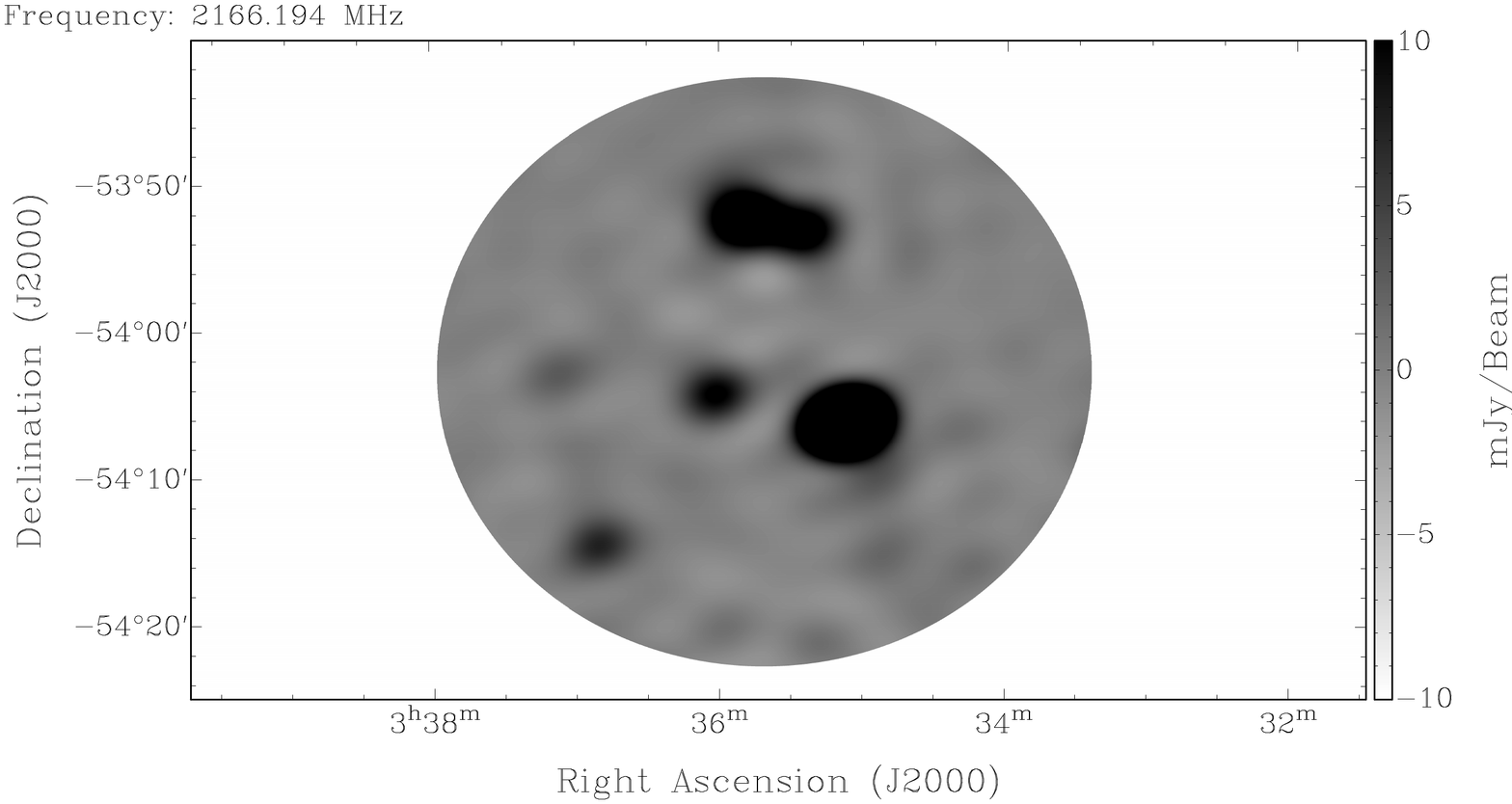}
 \hspace{-0.9cm}
   \includegraphics[width=0.43\textwidth,angle=-90]{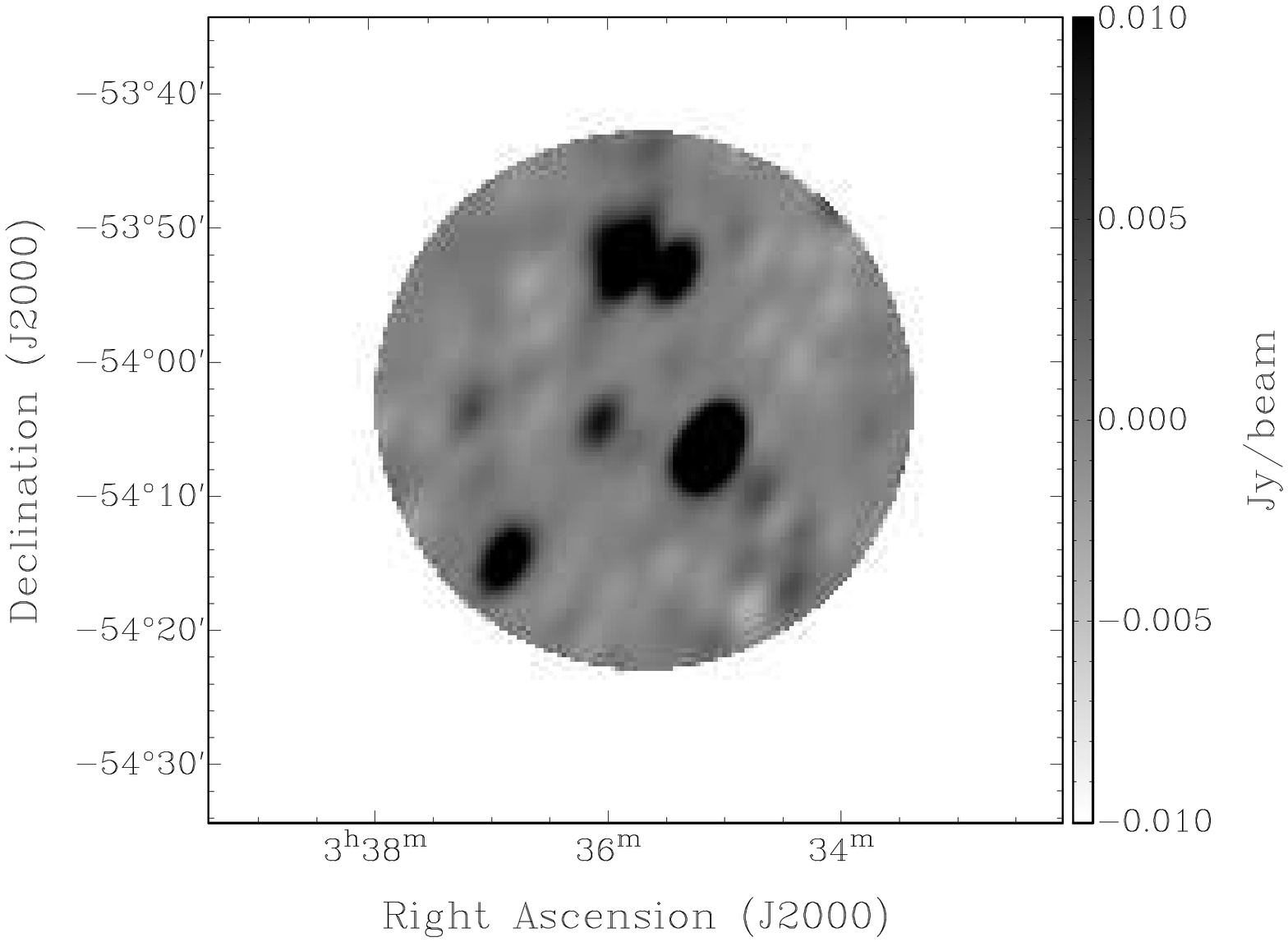}
\vspace{-1.cm}
\caption{Map of the ROI observed with ATCA at 2.1 GHz excluding the sixth antenna (left, $no6$ map) and observed with KAT-7 at 1.8 GHz (right). }
\label{fig:map_kat}
\end{figure}

The KAT-7 map can be directly compared to the $no6$ map, since the two images have a similar sinthesized beam and frequency range.
They are shown in Fig.~\ref{fig:map_kat}.
Given the large beam, the KAT-7 map is useful to study extended emission which is the reason behind the choice to perform these observations in addition to the ATCA ones.
In this way, we have two independent maps made by different telescopes, but with similar sensitivity and angular resolution, that can test the presence of a diffuse emission in RetII (which, we remind, is the main goal of the project).
On the other hand, for the KAT-7 maps, the source subtraction is not as well defined as for the ATCA $f15$ and $no6$ maps. Indeed, we have no KAT-7 long baseline to be employed to detect small-scale sources, so we have to rely on a model of sources detected with a different telescope (ATCA) at a different frequency and time.

For this reason, we find the KAT-7 data to typically bring less constraining information.
This holds true for all the main analyses described in the rest of the paper.
Nevertheless, the KAT-7 map provides independent consistency checks (as we will describe below) and therefore strengthen the robustness of the conclusions derived with the ATCA data.

\begin{figure}[t]
\centering
 \hspace{-0.8cm}
\includegraphics[width=0.54\textwidth]{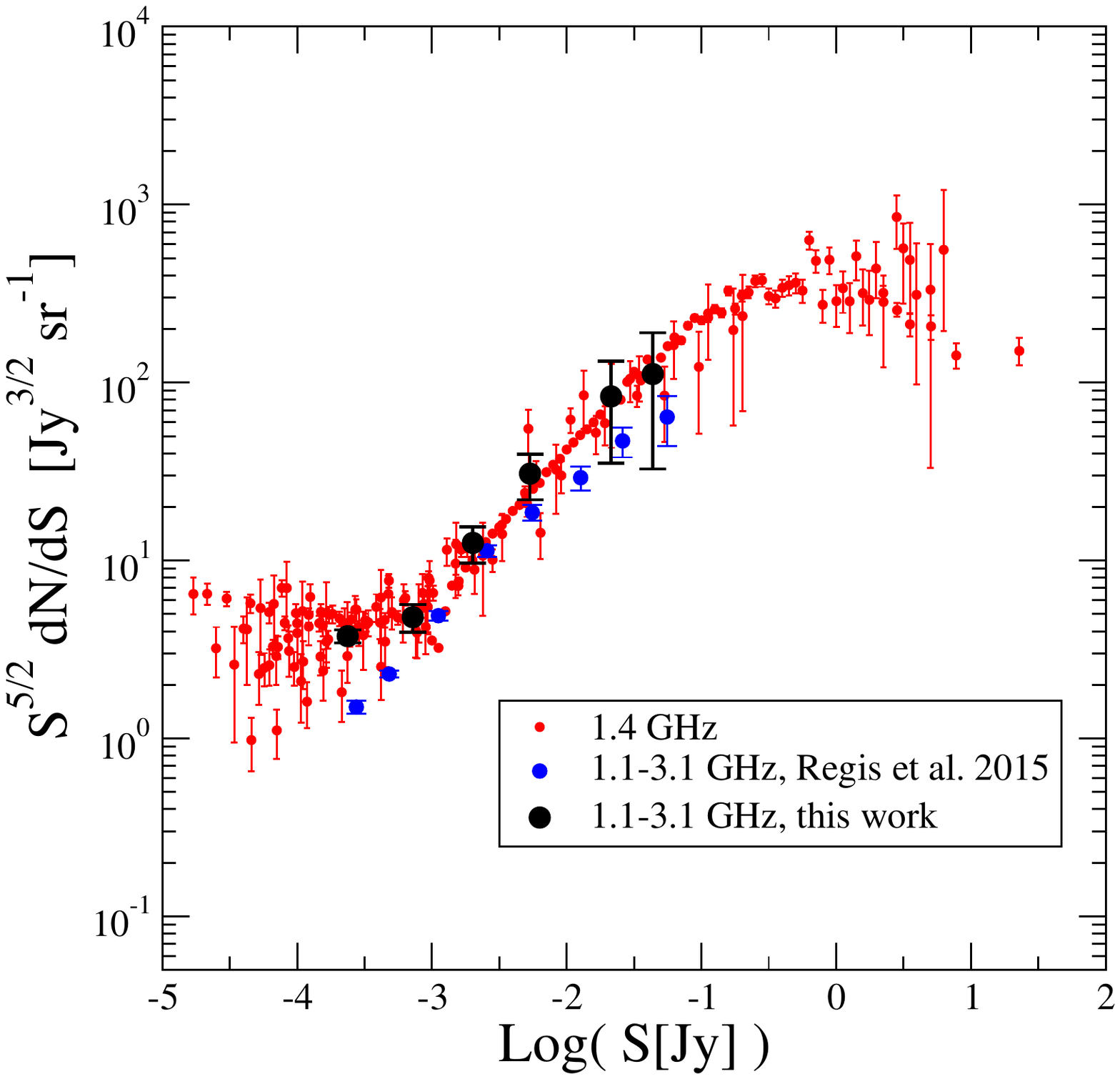}
 \hspace{-0.9cm}
\includegraphics[width=0.54\textwidth]{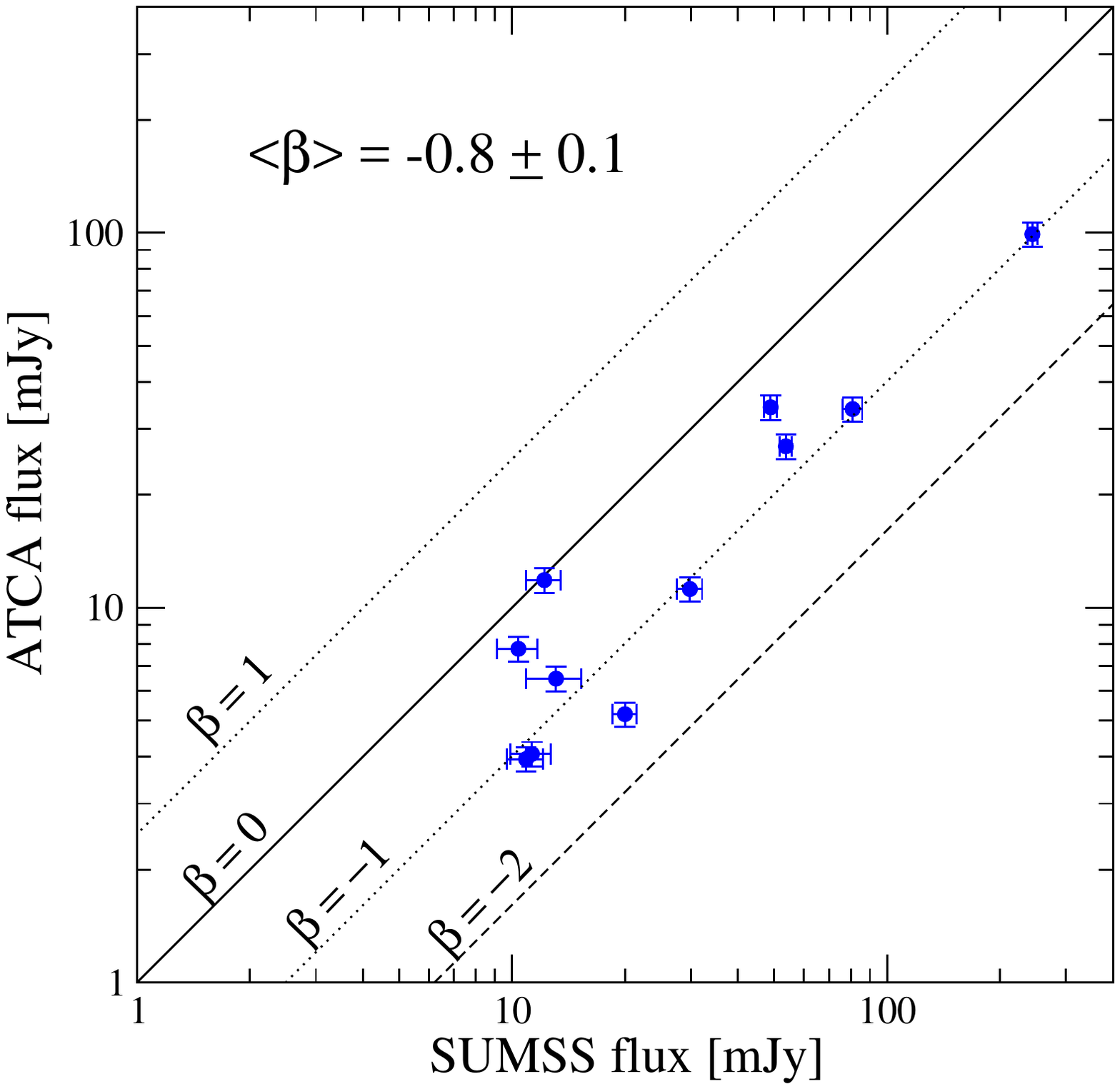}
\caption{Left: Source number counts in the ROI are shown with black dots, as a function of flux density. For comparison, the number counts of R15a (blue) and a compilation of observational data at 1.4 GHz from Ref.~\cite{DeZotti:2009an} (red) are reported. Right: Comparison of source flux densities in the SUMSS catalog with the results of this work. All the SUMSS sources in the ROI are matched. Lines show spectral index level with $\beta$ being $\beta=\ln (S_{ATCA}/S_{SUMSS})/\ln (2.1/0.843)$.   }
\label{fig:sources}
\end{figure}

\section{Source catalog}
\label{sec:cat}

We extracted sources employing the task SFIND of {\sc Miriad} on the $r_{-1}$ map and following the procedure described in R15a. We set the SFIND parameter $\alpha=0.1$, that roughly corresponds to a $5\sigma$ detection threshold.
To avoid strong primary beam effects we defined a ROI which includes only data within a radius of $20.1'$ (namely, 2$\sigma$ of the Gaussian approximation of the primary beam) from the center of the RetII pointing, see Fig.~\ref{fig:map_fullb}.
The catalog we derived has 285 entries corresponding to a total of 240 extracted sources with 21 cases being (possibly) multiple component sources.

The first ten entries of the catalog are reported in Table~\ref{tab:cat}.\footnote{The full catalog is available at \url{http://personalpages.to.infn.it/~regis/c3103.html}.}

In Fig.~\ref{fig:sources}a, we show the number counts of the extracted sources (black dots). The correction for incompleteness is computed as in R15a.

We notice an overproduction for the faint and bright ends with respect to the number counts derived in R15a (blue dots).
While the poor statistics of the bright end can explain the deviation (the last two bins include just a few sources and points are compatible within error bars), the mismatch of the first point is more worrisome. We ascribe it to two possible effects. In Fig. 9a of R15a a significant scatter among different fields has been found, suggesting that, for such a small ROI, the cosmic variance can play a relevant role. Then, the flux and counts derivations in R15a were made more awkward than here by the presence of a mosaic, which might have hidden possible systematic effects for faint sources. To fully address this issue, a simulation of the source reconstruction capability in R15a would be in order, but is clearly beyond the goal of this work.

Fig.~\ref{fig:sources}a shows that the number counts derived from the present catalog are in broad agreement with the literature (red points).

\begin{table*}
{\footnotesize
\centering
\caption{First ten lines of the catalog of sources extracted from the $r_{-1}$ map. In the column ``Multiple flag'', S stands for single source, while Mn refers to a component of the multiple source n.}
\label{tab:cat}
\begin{tabular}{rrrrrrcccccc}
\hline
\multicolumn{6}{c}{J2000}  &\multicolumn{2}{c}{Peak and Total flux density}& \multicolumn{2}{c}{Angular size} & P.A. & Multiple\\
\multicolumn{3}{c}{RA}  & \multicolumn{3}{c}{Dec} &$F^{peak}$ [mJy]&$F \pm \delta F$ [mJy]&$b_{maj}$ [$'$]&$b_{min}$ [$'$]&$\theta$ [deg]& flag \\
\hline

3 & 37 & 56.2 & -54 & 00 & 32.1 &  0.11  &  0.13 $\pm$ 0.02&  8.1 &  2.1 &  0.6 &  S\\
3 & 37 & 54.2 & -54 & 02 & 04.4 &  0.12  &  0.16 $\pm$ 0.02&  7.7 &  2.6 & 11.4 &  S\\
3 & 37 & 53.3 & -54 & 01 & 34.3 &  0.51  &  0.60 $\pm$ 0.04&  7.4 &  2.4 & -2.2 &  S\\
3 & 37 & 52.9 & -53 & 58 & 29.1 &  0.14  &  0.13 $\pm$ 0.02&  5.9 &  2.4 & -1.9 &  S\\
3 & 37 & 52.7 & -53 & 59 & 31.5 &  0.11  &  0.13 $\pm$ 0.02&  8.0 &  2.2 &  1.9 &  M1\\
3 & 37 & 51.5 & -53 & 59 & 28.7 &  0.14  &  0.19 $\pm$ 0.02& 10.2 &  2.1 & -3.9 &  M1\\
3 & 37 & 52.1 & -54 & 08 & 34.4 &  0.20  &  0.19 $\pm$ 0.03&  5.9 &  2.3 &  6.6 &  S\\
3 & 37 & 50.8 & -54 & 08 & 38.8 &  0.14  &  0.14 $\pm$ 0.03&  6.5 &  2.2 &  0.8 &  S\\
3 & 37 & 49.8 & -53 & 57 & 12.8 &  0.12  &  0.12 $\pm$ 0.02&  6.0 &  2.3 &  4.0 &  S\\
3 & 37 & 49.8 & -54 & 01 & 46.1 &  0.11  &  0.17 $\pm$ 0.02& 10.5 &  2.2 &  1.9 &  S\\
\hline
\end{tabular}
}
\end{table*}

\subsection{Comparison with SUMSS and KAT-7}
\label{sec:sumss}

We now compare our findings with the SUMSS radio catalog~\citep{Mauch:2003zh}.
In the considered ROI, SUMSS detected 11 sources at 0.843 GHz.
All of them have a counterpart in our catalog, with an average offset for the source positions of 3 arcsec. The latter is larger than our positional uncertainty (below 1 arcsec) and mainly given by SUMSS errors, due to their larger synthesized beam (about $45''$).

In Fig.~\ref{fig:sources}b, we compare SUMSS flux at 0.843 GHz with the flux derived in this work at 2.1 GHz.
In the case of sources with multiple components, we add up the flux densities of the various components.

The average spectral index is $\langle \beta \rangle=-0.8\pm 0.1$, which is a typical value for synchrotron radio continuum sources.
No significant outliers are present in the plot.
The SUMSS sources correspond to the the brightest (and largest) sources in our catalog.
We mentioned above that these are the sources that could present biggest issues in our imaging. The matching of position and flux with the SUMSS catalog, on the other hand, seem to exclude any significant problem.

Summarizing, from the two panels of Fig.~\ref{fig:sources}, we can conclude that our catalog is consistent with expectations and findings of previous surveys.

In the map built using data from the KAT-7 telescope, we extracted 10 sources.
Eight of them match with ten SUMSS sources (one of the KAT-7 sources is including three SUMSS sources because of the larger beam).
Only one SUMSS source is below threshold in KAT-7. This is consistent with the fact that the SUMSS catalog is complete to approximately 8 mJy, while the $1\sigma$ confusion limit due to the KAT-7 beam is slightly larger than 1 mJy, which means the two maps have comparable sensitivities. 
The average spectral index is again $\langle \beta \rangle=-0.8\pm 0.1$, as for the ATCA-SUMSS case, adding consistency to the whole picture.

The two KAT-7 sources that are not present in the SUMSS catalog do match with sources in our catalog derived from the $r_{-1}$ ATCA map.
We thus conclude that also the KAT-7 map do not show major concerns and its outcome is consistent with the ATCA maps.

\section{Radio counterpart for a possible gamma-ray emission}
\label{sec:excess}

\begin{figure}[t]
\vspace{-3cm}
\centering
\includegraphics[width=0.48\textwidth]{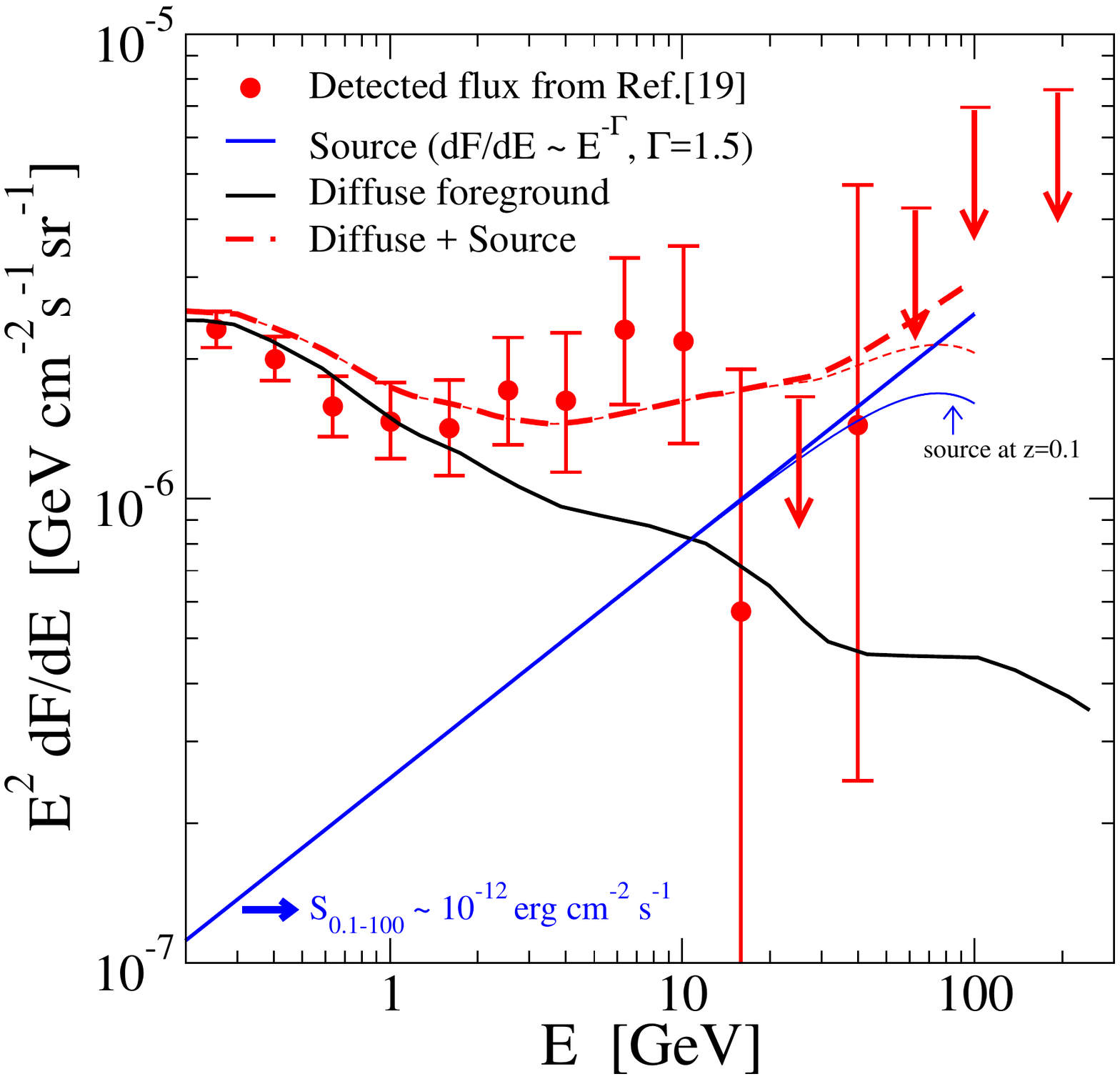}
\includegraphics[width=0.5\textwidth]{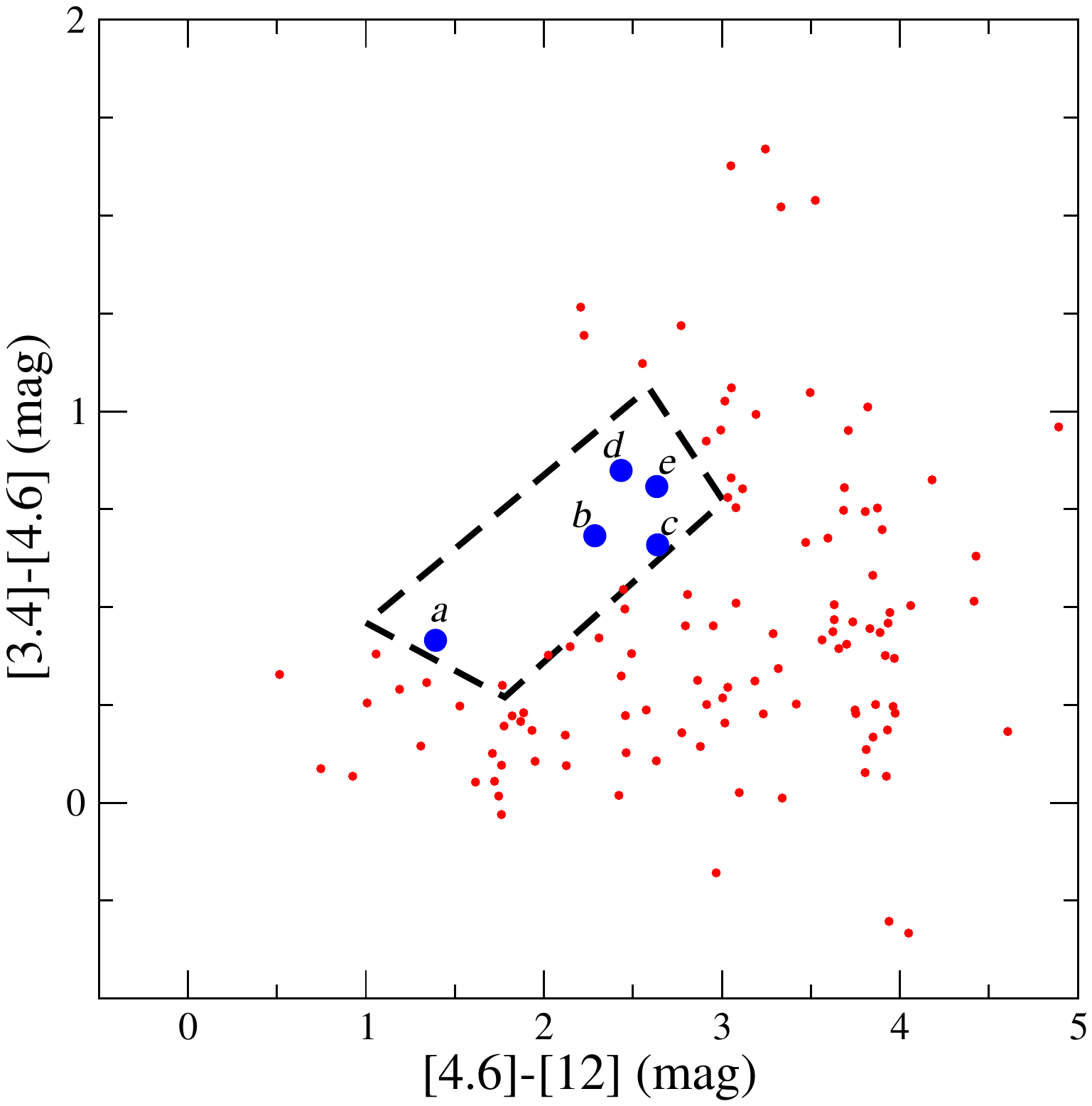}
\caption{Left: We show the energy spectrum of the \g-ray excess reported in Ref.~\cite{Geringer-Sameth:2015lua} (red points), together with the expected diffuse foreground (black line, from Ref.~\cite{Geringer-Sameth:2015lua}) and a simple power-law description of the source (blue lines). The sum of the two components is shown with dashed red lines and a case including absorption for a source located at $z=0.1$ is reported using thin lines. Right: Mid infrared color-color diagram of all sources from the catalog presented in Section~\ref{sec:cat} having a WISE counterpart within 3 arcsec. The region where BL Lacs are expected is shown with a black dashed contour taken from Ref.~\cite{Massaro:2017wws}. The properties of the sources corresponding to the blue points are reported in Table~\ref{tab:sources}.}
\label{fig:excess}
\end{figure}

As mentioned in the Introduction, a possible \g-ray emission from the direction of RetII was suggested in Ref.~\cite{Geringer-Sameth:2015lua}.
Even though the detection has still to be confirmed~\cite{Fermi-LAT:2016uux}, this potential signal can be very intriguing and has triggered a significant interest in the community.
In this Section, we investigate the possibility that one (or more) of the background sources detected in our catalog is the radio counterpart of the \g-ray emitter responsible for the signal.

In Fig.~\ref{fig:excess}a, we report the \g-ray spectrum derived in Ref.~\cite{Geringer-Sameth:2015lua} (red points). We show also the diffuse foreground expected for that region, again from Ref.~\cite{Geringer-Sameth:2015lua}. In order to fit the spectrum, a hard \g-ray source is required. For the sake of simplicity we assume a single power-law $E^{-\Gamma}$ for the flux energy spectrum and find that $\Gamma\sim 1.5$ is required (thick blue line). In Fig.~\ref{fig:excess}a, we show also that for an object at not-too-high redshift, the effect of absorption due to the extragalactic background light does not affect this simple description (thin blue line).
By integrating the differential flux in the range 0.1-100 GeV, we find that this putative source should have a \g-ray energy flux in the ball-park of $10^{-12}\,{\rm erg\,cm^{-2} \,s^{-1}}$.

We note that $\Gamma\sim 1.5$ means a quite hard spectrum, harder than the average of AGNs. Among the extragalactic \g-ray sources detected by the Fermi-LAT telescope, such low photon index is expected from blazars, and in particular, most likely, in BL Lacs, in the class termed high-synchrotron peaked (HSP)~\cite{Ackermann:2015yfk}. 

HSP are found to show a strong radio-\g-ray correlation~\cite{Ackermann:2011bg} (actually, the strongest among different classes of AGNs).
A \g-ray energy flux of $\sim 10^{-12}\,{\rm erg\,cm^{-2} \,s^{-1}}$ can correspond, using the empirical radio-\g-ray connection of Ref.~\cite{Ackermann:2015yfk}, to a radio flux density approximately in the range 0.1-10 mJy at 1.4 GHz (see Fig.~25 of Ref.~\cite{Ackermann:2015yfk}).

Therefore, even though the considerations we have done on the possible \g-ray source have no very strong statistical ground, they suggest that, if the source is indeed real and is an extragalactic background source, it should have a radio counterpart in the catalog presented in Section~\ref{sec:cat} (while, on the other hand, it is unlikely that the source is present in the SUMSS catalog which has a detection threshold of $\sim 8$ mJy).

Blazars show also a strong infrared-\g-ray connection \cite{Massaro:2012dh,Massaro:2016mgw}.
To identify possible BL Lac candidates, we perform a positional cross-matching of the sources extracted in Section~\ref{sec:cat} with the catalog of the NASA Wide-field Infrared Survey Explorer (WISE)~\cite{Wright:2010qw}, in order to search for infrared counterparts.
We restrict the positional mismatch to a maximum of three arcsec, taking into account the uncertainties of the two samples.
All the matching sources are shown with points in the mid-infrared color-color diagram of Fig.~\ref{fig:excess}b.

Blazars are found to occupy a well defined strip in this plane, separated from other infrared emitters~\cite{Massaro:2012dh,Massaro:2016mgw}. More in details, one can define a polygon (black dashed line in Fig.~\ref{fig:excess}b) where we expect to have only BL Lacs, with a contamination from other types of blazars (with softer spectrum) lower than 5\%~\cite{Massaro:2017wws}. However, our sample can still suffer of a larger contamination due to radio galaxies.
To identify possible HSP, we individually inspect all the sources falling well within the polygon. They are five, highlighted with thick blue points, and with properties reported in Table~\ref{tab:sources}. For all these sources, the positional mismatch between WISE and our catalog is below 0.5 arcsec, meaning that the probability of wrong cross-identification is very low.

The radio counterparts of \g-ray blazars appear to have quite flat radio spectrum, with $\beta\gtrsim -0.5$~\cite{Massaro:2013vpa}. In our inspection, we consider the energy spectrum, by comparing with the SUMSS catalog, as well as morphological properties of each source.  

\begin{figure}[t]
\vspace{-2.7cm}
   \centering
 \hspace{-0.6cm}
   \includegraphics[width=0.51\textwidth]{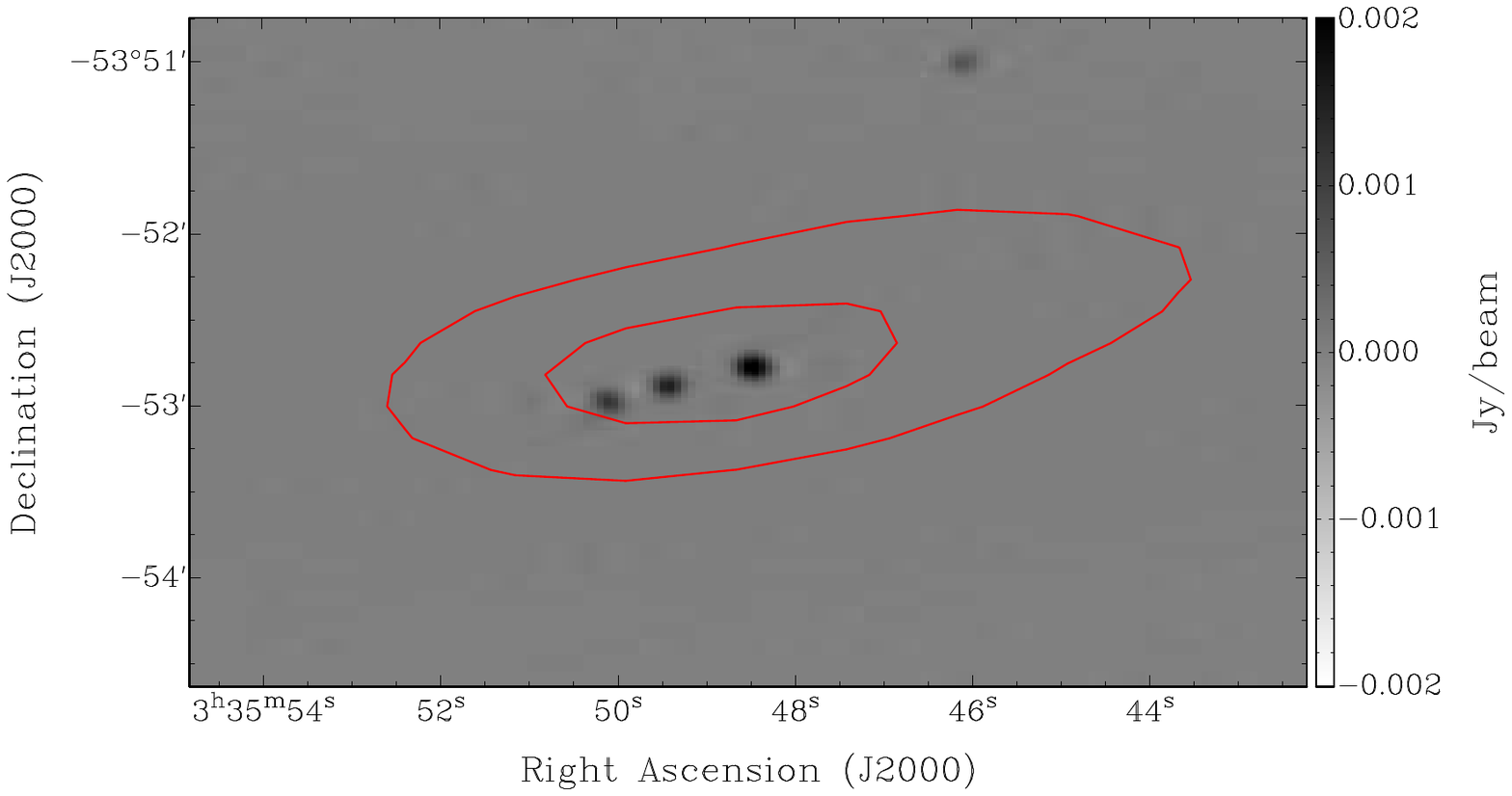}
 \hspace{-0.9cm}
   \includegraphics[width=0.54\textwidth]{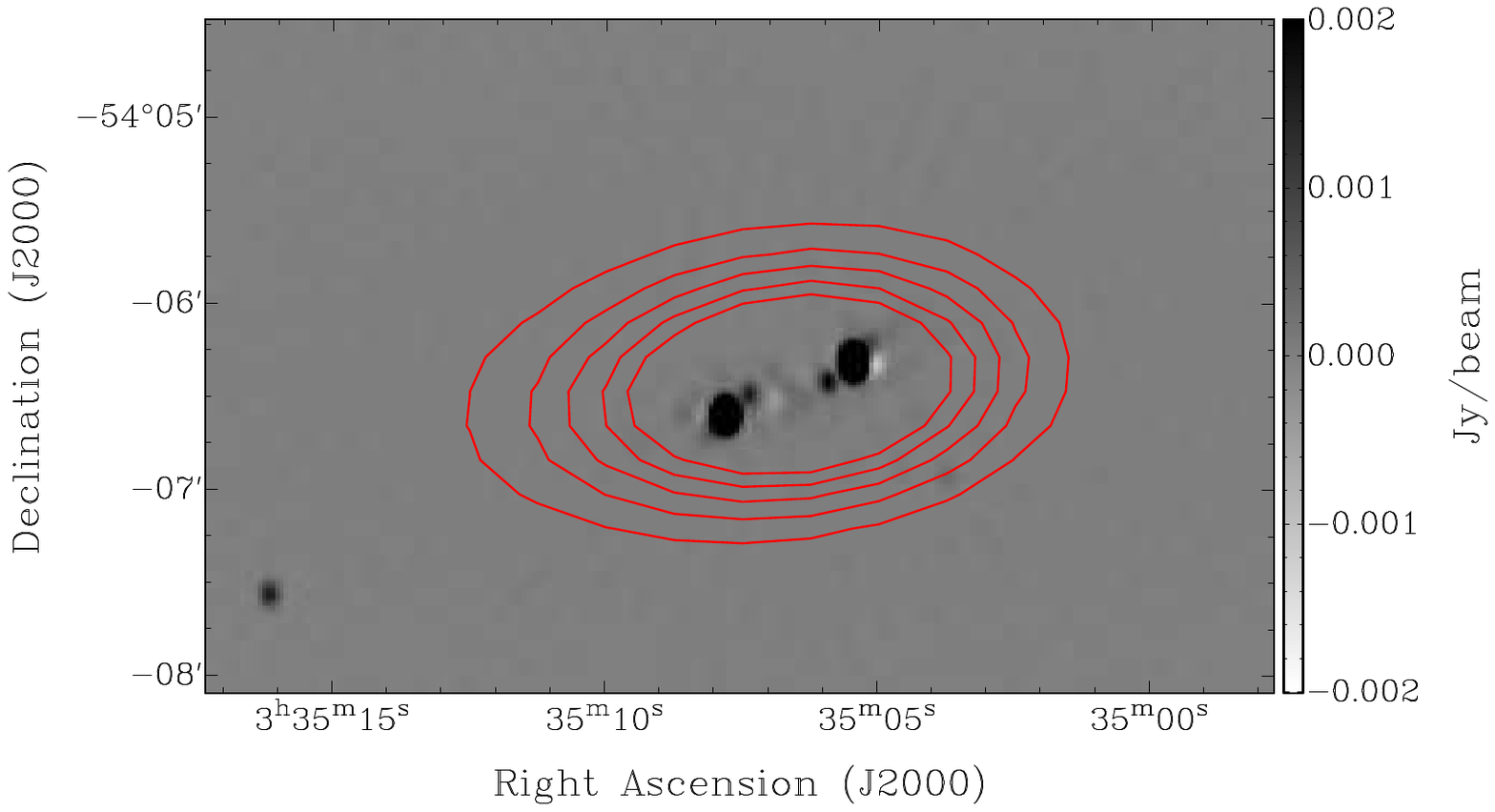}
\vspace{-4.cm}
\caption{Images of sources $c$ (left, see Table~\ref{tab:sources}) and PMN J0335-5406 (right) from the $r_{-1}$ map. SUMSS contours are overlaid in red, with steps of 20 mJy.}
\label{fig:sourceszoom}
\end{figure}

Source $a$ in Fig.~\ref{fig:excess}b has no companion in the SUMSS catalog but it seems to be present in the SUMSS map above the noise, with a flux just below the detection threshold of the SUMSS catalog. With a simple Gaussian fit of the source, we estimated a flux of $\simeq 6.1$ mJy from the SUMSS map, which would imply $\beta\simeq-1.2$.
The source has a probable optical counterpart (detected at angular separation of 0.3 arcsec with B-mag of 20.0 in Ref.~\cite{MRASS:2003} and at 0.4 arcsec and with B-mag of 19.7 in Ref.~\cite{Simon:2015fdw}), for which unfortunately we found no spectrum publicly available.
Given the low spectral index and that, in our catalog, this source was classified as part of a source with two components, we conclude that it is most likely not a blazar.
Source $b$ is instead point-like and with no apparent flux peak in the SUMSS map.
Source $c$ has a companion in the SUMSS catalogue ($\beta=-0.95$) and a structure well-compatible with a radio galaxy. Indeed, three components, that can be interpreted as a core and two lobes, are present in our catalog, and are shown in Fig.~\ref{fig:sourceszoom}a. This source is detected also with KAT-7 observations described above and in the Parkes-MIT-NRAO catalog~\cite{PMN:1994}.
On the other hand, these two observations have poor angular resolution and do include other sources in the beam. Therefore the spectral index we can derive from them is not  fully reliable. We also found an optical diffuse counterpart~\cite{Simon:2015fdw} at 0.3 arcsec with R-mag of 20.05 and SED typical of a non-thermal source.
This source is the brightest in the sample of Table~\ref{tab:sources}, with a total flux of $30.9\pm 1.6$ mJy. Therefore even though it is not a blazar and likely have a softer \g-ray spectrum with respect to what would be needed, it could nevertheless provide a non-negligible \g-ray contribution.
The case of source $d$ is similar to source $a$ with no companion in the SUMSS catalogue but with a clear peak above the noise in the SUMSS map. The estimated flux at 0.843 GHz is $\simeq 6.2$ mJy with $\beta\simeq-2.5$ which suggests it is not a blazar.
Source $e$ is point-like and with no apparent flux peak in the SUMSS map.
The absence of a corresponding peak in the SUMSS map for sources $b$ and $e$ implies their spectral index to be $\beta\gtrsim-1$.

We conclude that sources $b$ and $e$ do pass all our selection criteria and can be considered as compelling BL Lac candidates.
We also performed a search on the raw X-ray data taken by the SWIFT telescope~\cite{Burrows:2005gfa} in a follow-up of RetII in July 2016. The search has been however inconclusive. Indeed, source $b$ is slightly out of the SWIFT field of view, while a few counts have been observed in the region around source $e$, but well within the noise. The observations are too shallow to establish the presence of a counterpart (this holds true also for the other candidates in Table~\ref{tab:sources}).

We note that the angular separation of source $b$ from the center of RetII is larger than the best point-spread-function of the Fermi-LAT telescope ($\sim 0.1^\circ$ at high-energy).
Indeed, 94\% of the blazars identified by the Fermi-LAT in the 3FGL catalog have a counterpart within 6 arcmin~\cite{Acero:2015hja}. Therefore, source $b$ could be in principle disentangled from RetII. On the other hand, source $e$ is closer to the center and can be hardly distinguished from RetII by the Fermi-LAT itself. 

Using the empirical best-fit of the radio-\g-ray relation of Fig.~25 in Ref.~\cite{Ackermann:2015yfk}, we can translate the radio flux density of Table~\ref{tab:sources} into a corresponding \g-ray flux of $1.3\times 10^{-12}\,{\rm erg\,cm^{-2} \,s^{-1}}$ for source $b$ and $1.5\times 10^{-12}\,{\rm erg\,cm^{-2} \,s^{-1}}$ for source $e$.\footnote{We used $S_{0.1-100 {\rm GeV}}=1.6\cdot 10^{-12}\,(F_{\rm radio}/{\rm mJy})^{0.34}\,{\rm erg\,cm^{-2} \,s^{-1}}$. We cannot derive the uncertainty associated to these estimates from Ref.~\cite{Ackermann:2015yfk} since they report only the error on the slope, but not on the normalization of the radio-\g-ray relation.}
Remarkably, they are in the ball-park $\sim 10^{-12}\,{\rm erg\,cm^{-2} \,s^{-1}}$ that we identified above in order to explain the possible excess suggested in Ref.~\cite{Geringer-Sameth:2015lua}.

A spectroscopic optical follow-up of these two sources would be probably the easiest way to confirm their nature.

Let us also mention that, even though we employed the infrared cross-matching as our guideline, it is possible that other radio sources, not falling in the polygon of Fig.~\ref{fig:excess}b, can have bright \g-ray counterparts.
However, without such handles, the search would become rather awkward, and a detailed multi-wavelength classification of all the sources in our catalog is clearly beyond the goal of this work.

Finally, we note that the source PMN J0335-5406 discussed in Ref.~\cite{Fermi-LAT:2016uux} as radio counterpart of the possible \g-ray emission (but with this interpretation somewhat disfavored by optical/near-infrared data~\cite{Fermi-LAT:2016uux}) is found in our analysis with a relatively low spectral index ($\beta=-1.0\pm 0.1$) and is associated to a multiple component source with morphological properties typical of a misaligned AGN, as shown in Fig.~\ref{fig:sourceszoom}b.

\begin{table}
\centering
\begin{tabular}{|c|rrr|rrr|c|c|}
\hline
id &\multicolumn{6}{|c|}{J2000}& Distance  &  Flux density     \\
&\multicolumn{3}{|c}{RA} & \multicolumn{3}{c|}{DEC} & arcmin  & mJy \\
\hline
$a^*$ & 03 & 34 & 46.0 & -54 & 08 & 43.7 & 9.9  & $1.26 \pm 0.07$ \\
$b$   & 03 & 34 & 29.6 & -53 & 54 & 44.4 & 13.3 & $0.59 \pm 0.04$ \\
$c^*$ & 03 & 35 & 49.4 & -53 & 52 & 52.6 & 10.2 & $2.39 \pm 0.15$ \\
$d$   & 03 & 34 & 52.5 & -53 & 50 & 19.4 & 14.5 & $0.53 \pm 0.04$ \\ 
$e$   & 03 & 35 & 32.4 & -54 & 04 & 12.1 & 1.7  & $0.89 \pm 0.05$ \\ 
\hline
\end{tabular}
\caption{Properties of the objects included in the box Fig.~\ref{fig:excess}b. The most promising BL Lac candidates are the targets $b$ and $e$.\hspace{\textwidth}
{\footnotesize $^*$Objects $a$ and $c$ are parts of multiple component sources: the quoted properties refer to the radio component that is closer to the infrared source. The flux of the whole source is $2.1\pm 0.1$ mJy for $a$ and $30.9\pm 1.6$ mJy for $c$.}}
\label{tab:sources}
\end{table}

\section{Diffuse emission}
\label{sec:diffuse}

\begin{figure}[t]
\vspace{-2.5cm}
   \centering
 \hspace{-0.8cm}
   \includegraphics[width=0.54\textwidth]{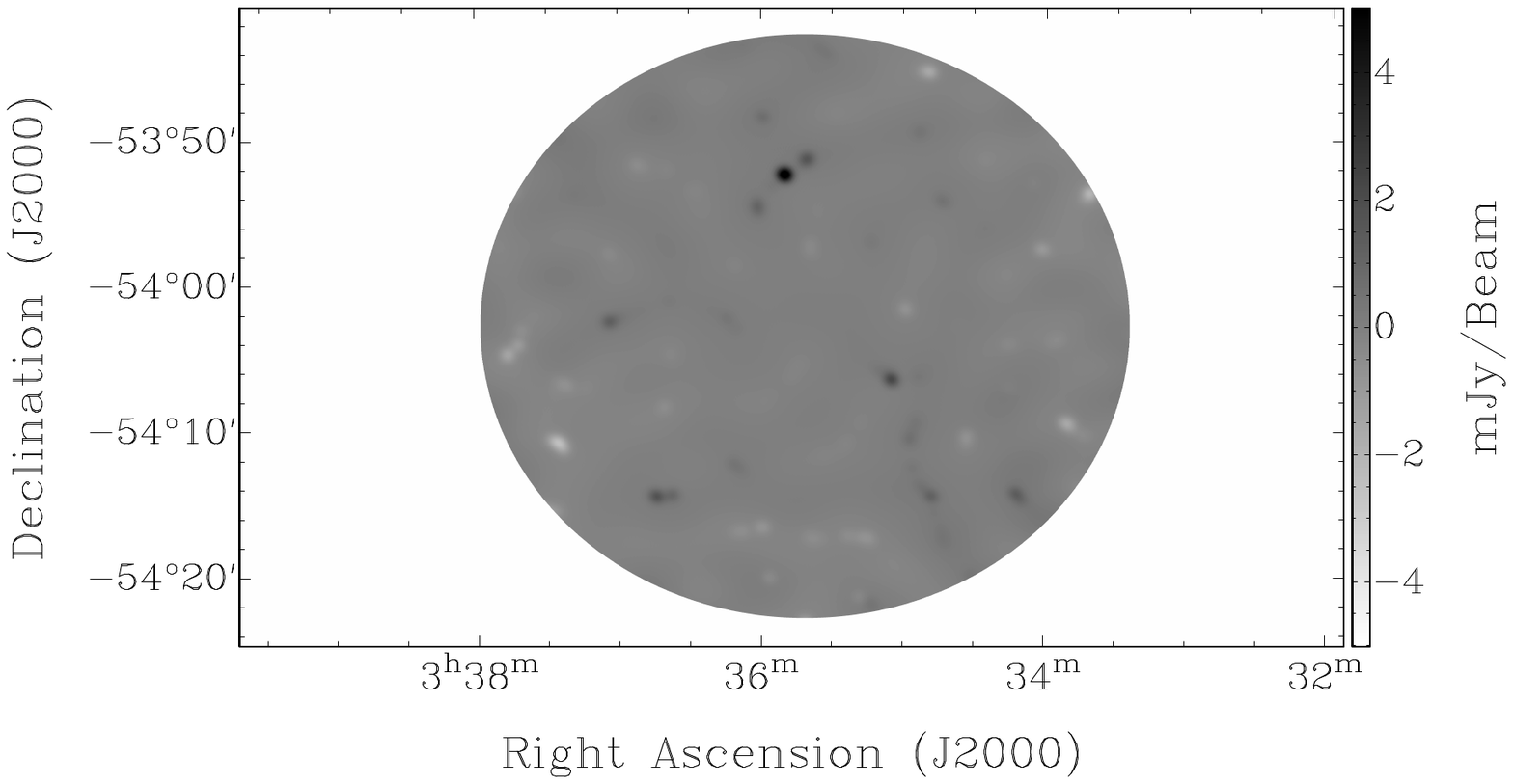}
 \hspace{-0.9cm}
   \includegraphics[width=0.54\textwidth]{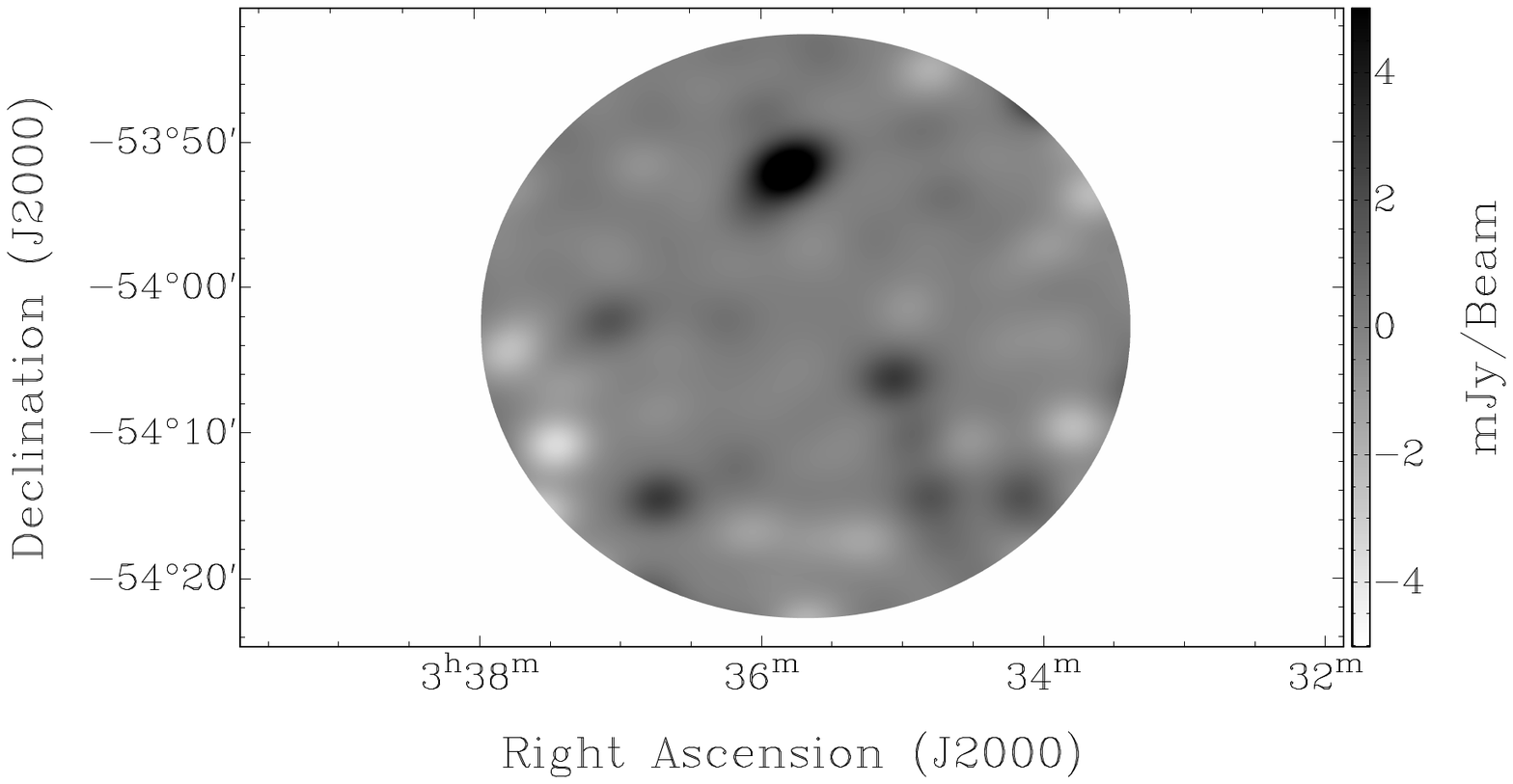}
\vspace{-4.5cm}
\caption{ {\bf Source subtraction}. Map obtained after subtracting sources in the UV-plane.
In the left panel, the map is tapered with FWHM=$15''$, while, in the right panel, the image is obtained removing baselines involving the sixth antenna. }
\label{fig:map_sub}
\end{figure}

To test the presence of a diffuse emission from RetII \footnote{Here, diffuse means scales $\gtrsim 1'$.}, we consider the $f15$ and $no6$ maps.
We first subtract small-scale sources detected in Section~\ref{sec:cat} using the $r_{-1}$ map and then compare the obtained maps with models, as detailed below.
The maps of the rms noise after source subtraction are again computed with the SEXTRACTOR package as done for the rms of the $r_{-1}$ map and described in R15b.

The $\sigma_{rms}^i$ of each pixel $i$ enters in the Gaussian likelihood defined by:
\be 
\mathcal{L}=e^{-\chi^2/2} \;\;\; {\rm with} \;\;\; \chi^2=\frac{1}{N_{pix}^{beam}}\sum_{i=1}^{N_{pix}} \left(\frac{S_{th}^i-S_{obs}^i}{\sigma_{rms}^i}\right)^2\;,
\label{eq:like}
\ee
where $S_{th}^i$ is the theoretical estimate for the brightness (see Section~\ref{sec:mod}), $S_{obs}^i$ is the observed brightness (see Section~\ref{sec:sub}), $N_{pix}$ is the total number of pixels in the ROI, and $N_{pix}^{beam}$ is the number of pixels in a synthesized beam.

\subsection{Subtraction of small-scale sources}
\label{sec:sub}
Small-scale discrete sources are characterized by means of the $r_{-1}$ map as described above. 
The detected structures vary from few to few tens of arcsec. 
In order to reduce the confusion noise of the $f15$ and $no6$ images, we subtract the sources in the visibility plane of the two maps. 
To this aim we employ the task UVMODEL in {\sc Miriad} and the CLEAN component of the $r_{-1}$ map as the input source model (see R15b).
The resulting visibilities are then reduced and imaged following the same pipeline as for the original maps.

Our observational setup, having a core array and a distant antenna, does not have intermediate baselines, and is basically not sensitive to scales in the range (approximately) from $10''$ to $1'$. As already mentioned, this is due to a ``hole" in the UV plane corresponding to the lack of baselines of length in between $\sim 100$ m (i.e. longest baseline of the core) and 4 km (i.e. baselines involving the sixth antenna). 

The median radio source angular size of extragalactic background objects with flux lower than about 100 mJy is below 10 arcsec, see, e.g., Ref.~\cite{Windhorst:1990}.
Clouds within the dSph or in the Galaxy might contribute at few tens of arcsec scales, but their presence is likely to be negligible.
Therefore, although having a complete coverage of the UV plane would be clearly ideal, our observing setup is adequate to study diffuse emission from RetII on the expected scales (above 1 arcmin) and also to characterize (and subtract) the vast majority of the background sources. 
On the other hand, as noted before, the catalog contains a handful of sources (the brightest ones) having sizes above a few arcsec. 
Their structure is recovered in a suboptimal way. This impacts on their subtraction and can affect the rms of the source subtracted maps.

\begin{figure}[t]
\vspace{-3cm}
\centering
   \includegraphics[width=0.49\textwidth]{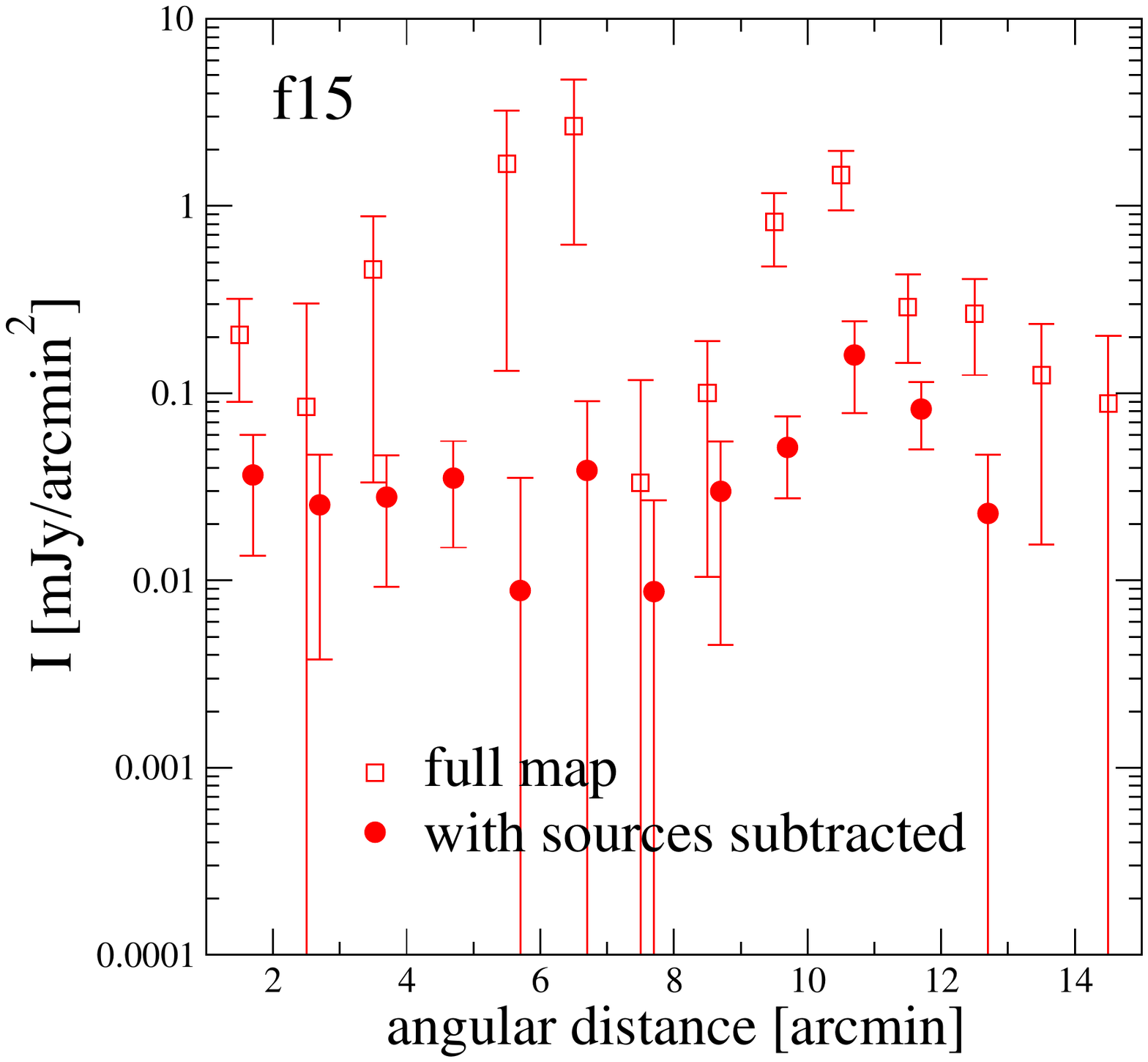}
   \includegraphics[width=0.49\textwidth]{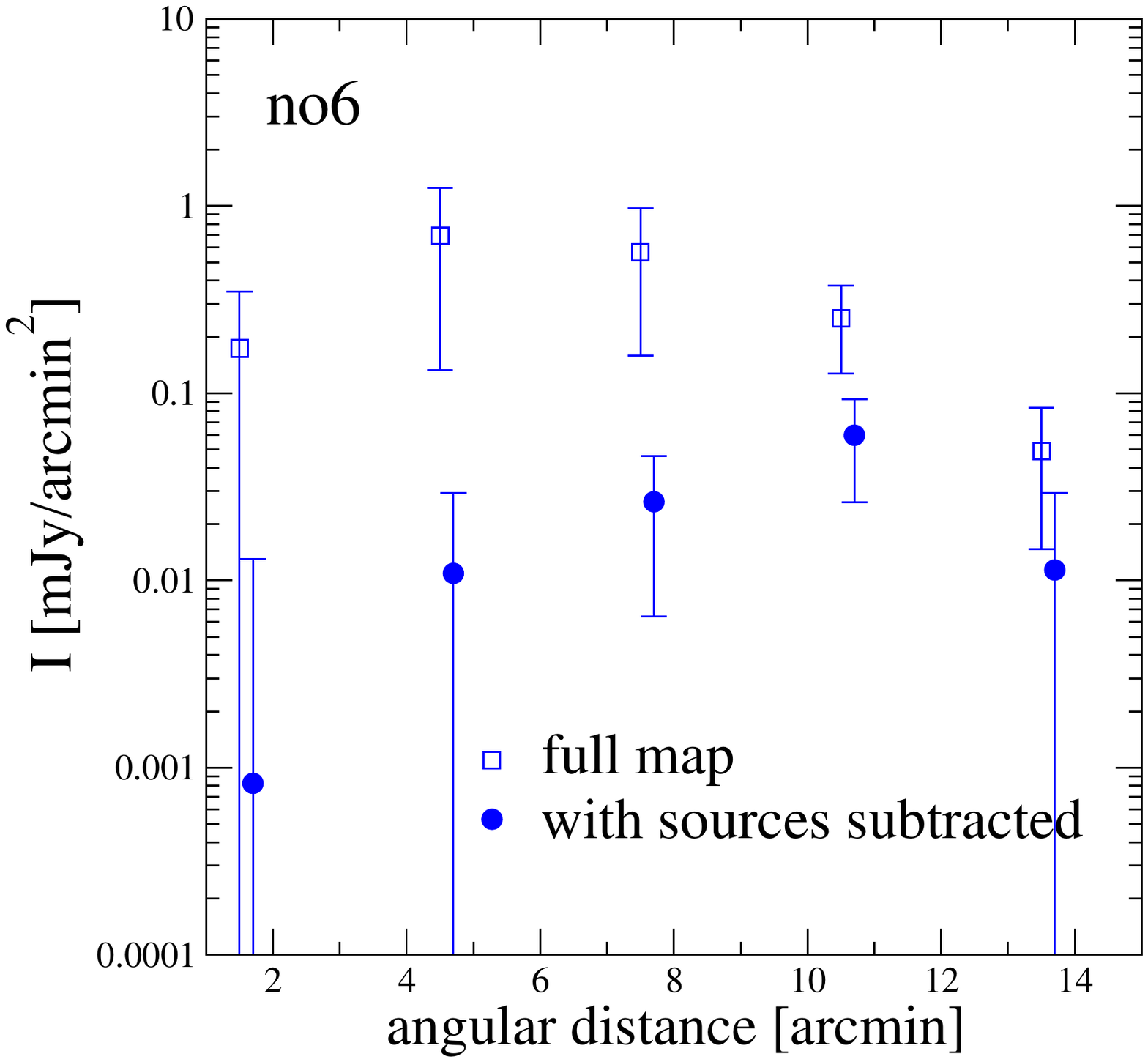}
\caption{ {\bf Radial profiles}. Measured emission averaged in spherical annuli in the $f15$ (left, width of annuli $=1'$) and $no6$ (right, width of annuli $=3'$) maps, as a function of the distance from the center. Empty squares refer to the original map, while filled circles show the emission after source subtraction.}
\label{fig:annuli}
\end{figure}

In Fig.~\ref{fig:annuli}, we show the radial distribution of the observed surface brightness in the $f15$ (left) and $no6$ (right) maps before (empty squares) and after (filled circles) source subtraction. 
The points are the average of the emission in spherical annuli of width of 1 arcmin for the $f15$ map and of 3 arcmin for the $no6$ map.
The error bars are computed by summing in quadrature the average of the rms maps and the standard deviation of the emission within each annulus.

Even though, for the reasons discussed above, some remnants of bright sources remain present in the subtracted maps (see also Fig.~\ref{fig:map_sub}), the source subtraction provides a significant improvement.
The average noise decreases from about 150 to 30 $\mu$Jy/beam in the $f15$ case and from about 450 to 160 $\mu$Jy/beam in the $no6$ case.
Note that the noise in the $no6$ map is more than a factor of 5 higher than in the $f15$ map. On the other hand, the beam is a factor of 16 larger. Therefore the $no6$ map can in principle be more sensitive to very extended emissions, while the $f15$ map is superior at intermediate scales. This is the reason why we investigate both maps.

The profiles in Fig.~\ref{fig:annuli} can be used to constrain any model of diffuse emission from RetII.
Note they show some deviation from zero, even for the maps after source subtraction. This is however due to residuals of discrete sources rather than to truly diffuse emission, as we will discuss in the next Section.

\subsection{Models}
\label{sec:mod}

The synchrotron diffuse emission in RetII can be computed by convolving the descriptions of the population of high-energy electrons and positrons and of the magnetic field in the dSph.

\begin{figure}[t]
\vspace{-3cm}
\centering
   \includegraphics[width=0.49\textwidth]{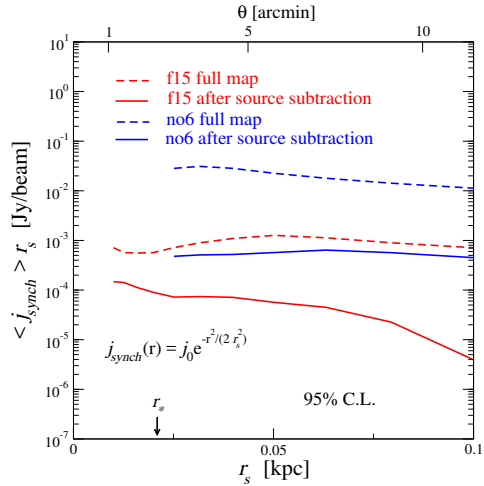}
\caption{ {\bf Bounds on emissivity}. 95\% C.L. observational upper limits on the emissivity $\langle j_{synch}\rangle=3\,r_s^{-3}\int^{r_s}_0 dr\,r^2 j_{synch}(r/r_s)$ versus spatial extension $r_s$, for the maps considered in our analysis. The spatial profile of the emissivity is modeled with a Gaussian law.}
\label{fig:j0}
\end{figure}

A simpler approach is, on the other hand, to introduce a phenomenological functional form for the synchrotron emissivity $j_{synch}(r)$. Once integrated along the line of sight and in the angular beam of the map, it provides the predicted flux density (see Eq.~2 in R15b).
We consider a Gaussian law with two free parameters, $j_{synch}(r)=j_0\,\exp[-r^2/(2\,r_s^2)]$, and derive bounds on the level and spatial size of the emission. 
Alternative functions used in the literature (as, e.g., $\beta$-models) would produce very similar results (see R15b).
In Fig.~\ref{fig:j0}, we show bounds on the quantity $\langle j_{synch}\rangle=3\,r_s^{-3}\int^{r_s}_0 dr\,r^2 j_{synch}(r/r_s)$ versus spatial extension $r_s$.
Note how they are significantly improved by the subtraction of sources, both for the $f15$ and $no6$ maps.

By evaluating the likelihood ratio between the null signal (i.e., $S_{th}=0$ in Eq.~\ref{eq:like}) and a model, we can also estimate the significance of detection of the latter.
If we consider a spatially flat term, we would get a detection at high confidence ($6.5\sigma$ in the $f15$ map and $4.7\sigma$ in the $no6$ map), as clear even by eye from Fig.~\ref{fig:annuli}.
On the other hand, by considering a more physical picture and limiting $r_s$ in the above Gaussian model to be $\lesssim r_*$, the $\chi^2$ difference significantly reduces,
meaning that a spatially constant term is highly preferred over a physical model. This happens at $4.5\sigma$ in the $f15$ map and $4.4\sigma$ in the $no6$ map.
We thus interpret the emission detected in the maps after source subtraction as a flat noise term (due to residuals of sources) rather than as an astrophysical signal from RetII.

Moving now to a more detailed description, we need to model the magnetic field and the population of of high-energy electrons and positrons.
The transport of $e^+-e^-$ in the dSph can be described as a diffusive process governed by the equation:
\be
 -\frac{1}{r^2}\frac{\partial}{\partial r}\left[r^2 D(r,p)\frac{\partial f_e}{\partial r} \right] 
  +\frac{1}{p^2}\frac{\partial}{\partial p}(\dot p p^2 f_e)=
  s_e( r, p)
\label{eq:transp}
\ee
where we assumed spherical symmetry and stationarity. In Eq.~\ref{eq:transp}, $r$ is the radius, $p$ is the momentum, $D(r,p)$ is the diffusion coefficient, $\dot p$ is the energy loss term due to radiative processes (we include synchrotron and inverse Compton on CMB), $f_e(r,p)$ is the equilibrium $e^+-e^-$ distribution function, and $s_e( r, p)$ is the source function.
Further details on the modeling and on the numerical solution of Eq.~\ref{eq:transp} can be found in the Appendix of R14b.

The only source of injection we will consider for the $e^+-e^-$ population is given by DM particles, through their annihilation or decay.
The DM source function is defined by means of $q_e(r,E)dE=4\pi \,p^2\,s_e(r,p)dp$, where $E$ is the energy of the injected particle. 
In the annihilating scenario, it takes the form:
\be
q^a_e(E,r)=\langle\sigma_a v\rangle\,\frac{\rho(r)^2}{2\,M_{\chi}^2} \times \frac{dN_e^a}{dE}(E) \;,
\label{eqQ}
\ee
where $\langle \sigma_a v\rangle$ is the velocity-averaged annihilation rate, $M_{\chi}$ is the mass of the DM particle, $\rho(r)$ is the halo mass density profile, and $dN_e^a/dE$ is the number of electrons/positrons emitted per annihilation in the energy interval $(E,E+dE)$.
\footnote{In the case of WIMP as a Dirac fermion, the overall factor becomes 1/4, while 1/2 is appropriate for the more common cases of WIMP as a boson or Majorana fermion.}
We assume a smooth (without substructures), spherically symmetric and static dark matter distribution.

In the decaying DM scenario, the source function is described by:
\be
q_e^d(r,E)=\Gamma_d\,\frac{\rho(r)}{M_{\chi}} \times \frac{dN_e^d}{dE}(E) \;,
\label{eqQ2}
\ee
where $\Gamma_d$ is the decay rate and $dN_e^d/dE(E)$ is the number of electrons/positrons emitted per decay in $(E,E+dE)$.

Following Ref.~\cite{Bonnivard:2015tta}, we employ the Einasto model for the DM spatial profile:
\be
\rho(r)=\rho_{-2}\exp\left\{-\frac{2}{\alpha}\left[\left(\frac{r}{r_{-2}}\right)^\alpha-1\right]\right\}
\ee 
As reference values for the three parameters entering the halo description, we use the peak of the marginalized posteriors of Ref.~\cite{Bonnivard:2015tta}: $\rho_{-2}=7\times 10^7\,M_\odot/{\rm kpc}^3$, $r_{-2}=0.2$ kpc and $\alpha=0.4$.
This choice provides a $J$-factor of $\log_{10}[J(\alpha_{int}=0.5^\circ)/({\rm GeV^2\,cm^{-5}})]=18.9$ which corresponds (by chance) to approximately the 1$\sigma$ lower limit found in Ref.~\cite{Bonnivard:2015tta}. Since this set of values corresponds to a conservative choice is suitable to be used to derive upper bounds on the annihilation/decay rate.

The magnetic properties of dSphs are poorly known. The extremely low content of gas and dust makes polarization measurements very challenging.
This fact can strongly affect the predictions for the synchrotron signal.
We will follow the approach described in Sec.~4.3 of R15b, where two physical arguments were suggested to have an handle on the magnetic field description.
The first argument relies on the extrapolation of an empirical scaling relation between the star formation rate (SFR) of a galaxy and its magnetic field strength, found to hold in the Local Group~\cite{Chyzy:2011sw}, and on assuming a typical value for the SFR in ultra-faint dSph (with current observational data, the SFR in RetII can only be estimated in a quite approximate way). The second argument concerns the magnetization of the medium surrounding galaxies, like the Milky Way, (due to, e.g., Galactic outflows), since observationally the strength of intergalactic magnetic fields appear to be larger than a fraction of $\mu$G~\cite{Kronberg:2007wa}.

Putting these arguments together we can assume a magnetic field maximal strength of $B_0=1\,\mu$G and describe the spatial shape by means of $B=B_0\,e^{-r/r_*}$.

Turbulence properties of the magnetic field are as well poorly known, from the observational point of view. 
We consider a diffusion coefficient with normalization and spectrum defined in order to have a Milky-Way like diffusion within the stellar region and then growing exponentially in the outskirt: $D=3\cdot 10^{28}\,(E/{GeV})^{0.3}\exp(r/r_*)\,{\rm cm^2/s}$.

\section{Constraints on the WIMP parameter space}
\label{sec:bound}

In the derivation of bounds on the WIMP annihilation/decay rate as a function of the DM mass, we assume a definite astrophysical scenario, described in the previous Section. For an estimate of the size of the uncertainties due to the modeling of the magnetic properties and of the DM halo profile, see the discussion in R14 on the ultra-faint dSph targets.

Concerning the particle DM model, we do not consider a specific theory beyond the Standard Model, but rather we select four possible benchmark final states of annihilation/decay: $b\bar b$, $W^+W^-$, $\tau^+\tau^-$ and $\mu^+\mu^-$.
For the $b\bar b$ channel, the $e^+-e^-$ particles are injected by means of production and decay of $\pi^\pm$. This is also the case for $W^+W^-$, that in addition has a non-negligible contribution from direct decay into $e^+-e^-$ at energies close to the DM mass. The final states $\tau^+\tau^-$ can again produce $\pi^\pm$ (through semi-hadronic decay) but can also decay in $\mu^+-\mu^-$ which in turn decay into $e^+-e^-$. The resulting energy spectra becomes harder as we move from hadronic to leptonic production.

\begin{figure}[t]
\vspace{-3cm}
\centering
   \includegraphics[width=0.49\textwidth]{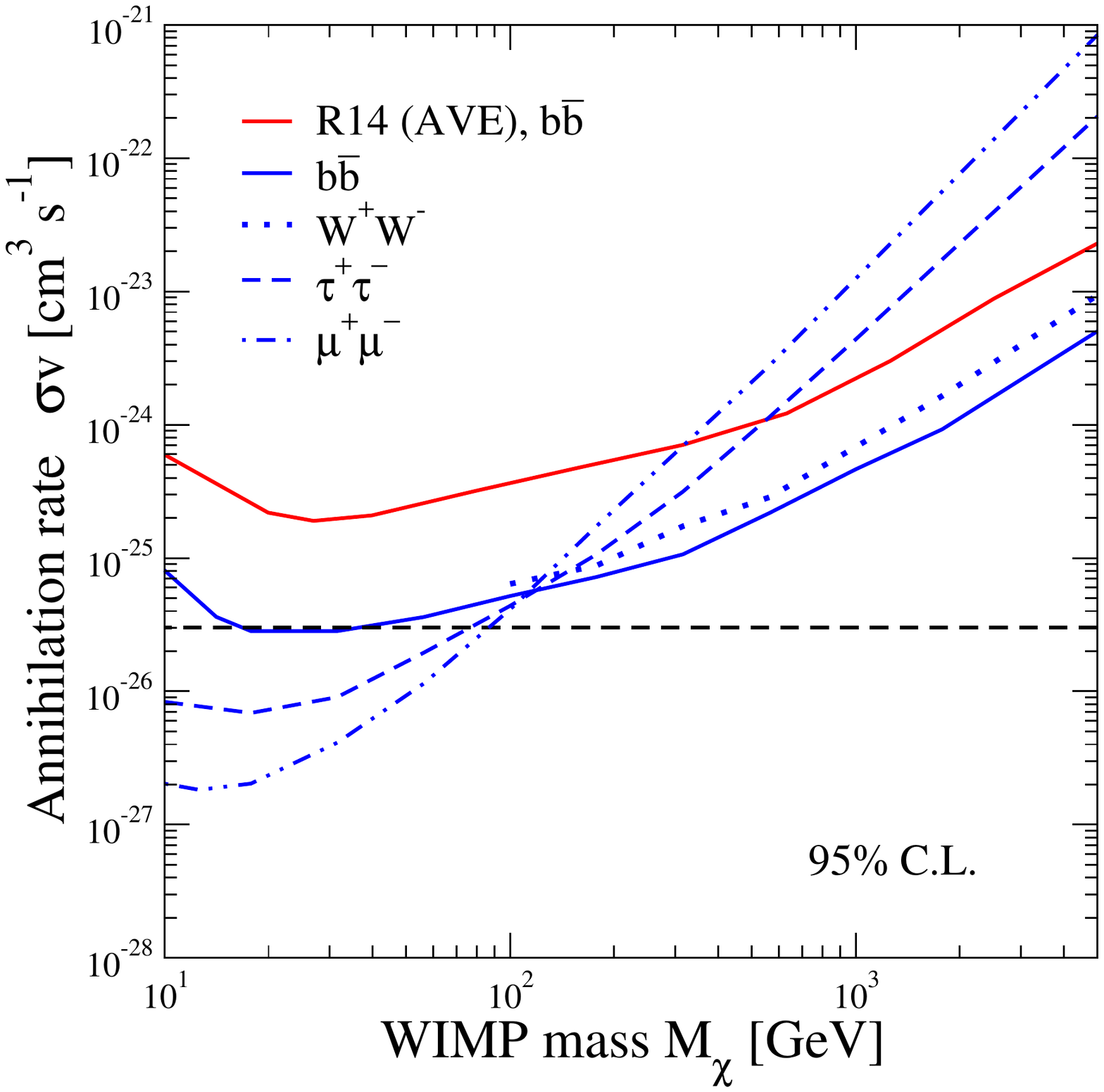}
   \includegraphics[width=0.49\textwidth]{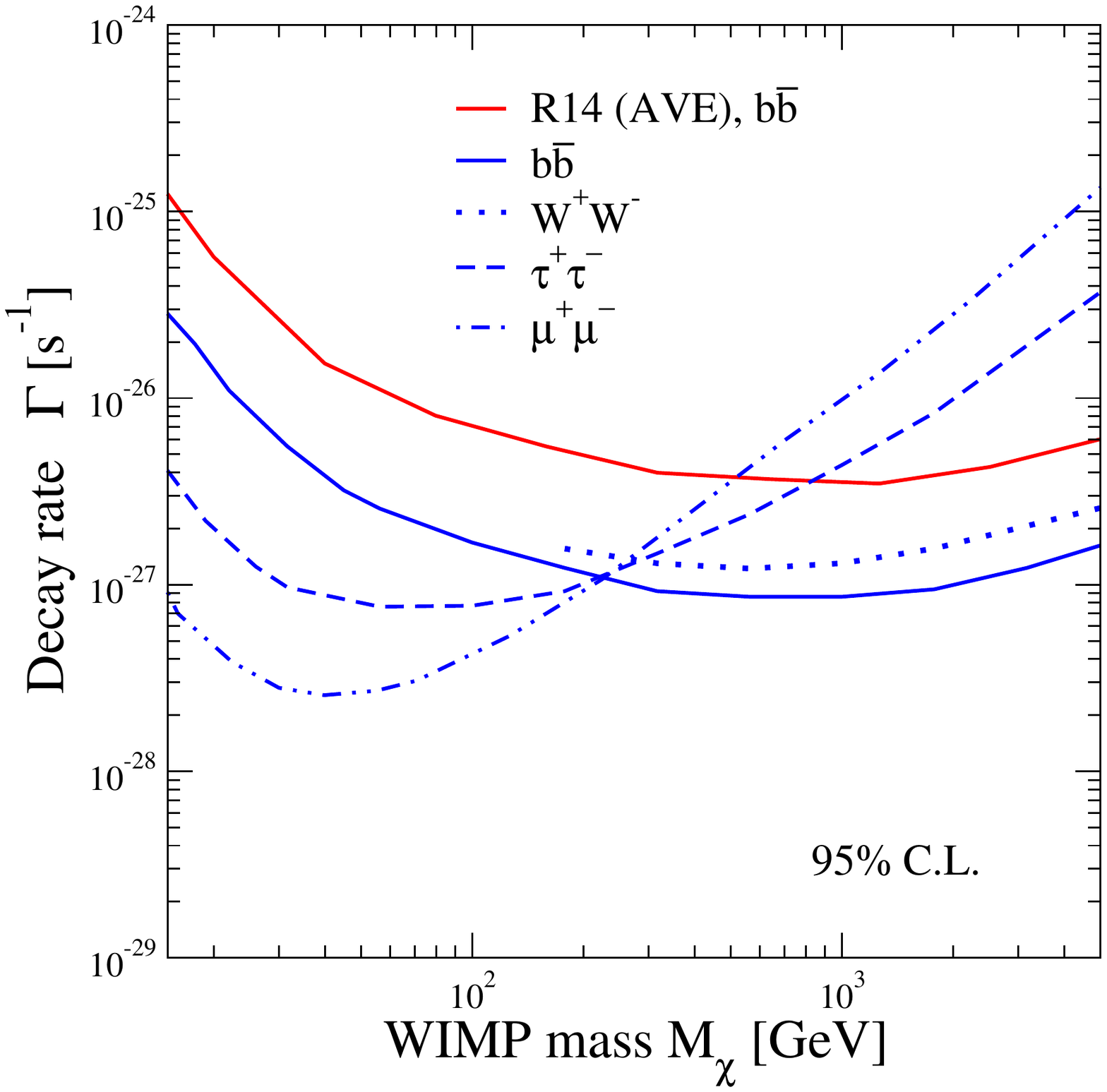}
\caption{ {\bf DM bounds}. Left panel: 95\% C.L. upper bounds on the velocity averaged annihilation cross-section as a function of the WIMP mass. We show the constraints for annihilation in the $b\bar b$ (solid), $W^+W^-$ (dotted), $\tau^+\tau^-$ (dashed) and $\mu^+\mu^-$ (dashed-dotted) channels, and, for comparison, the bounds derived in R14 (red, only for $b\bar b$). Right panel: 95\% C.L. upper bounds on the decay rate $\Gamma=1/\tau$ as a function of the DM mass, for the same final states as in the left panel.}
\label{fig:bounds}
\end{figure}

Bounds are shown in Fig.~\ref{fig:bounds}. They are derived taking the best limit between the ones derived from the $f15$ and $no6$ maps (as discussed above, the KAT-7 map provides weaker bounds). In practice, the $f15$ map results always more constraining.

The synchrotron power peak at 2 GHz and for a magnetic field of $1\,\mu$G corresponds to an $e^+-e^-$ energy of $\sim 20$ GeV.
The soft spectrum of the $b\bar b$ and  $W^+W^-$ final states implies that $e^+-e^-$ of 20 GeV are produced in a less efficient way with respect to $\tau^+\tau^-$ and $\mu^-\mu^-$, for DM masses below a few hundreds of GeV. This explain the difference in the bounds at low DM masses.
When the kinematic depletion is no longer acting, the higher multiplicity of the $b\bar b$ and  $W^+W^-$ channels makes the associated bounds stronger than in the leptonic case.

We note that, below 100 GeV, the thermal annihilation rate is strongly constrained in the leptonic channels, and is nearly approached for the $b\bar b$ final state (except at very low masses).

In Fig.~\ref{fig:bounds}, we show for comparison the bounds obtained in R14 (thin red line, only for the $b\bar b$ final state of annihilation, for the sake of brevity, and considering their AVE case that is very similar to the astrophysical scenario considered here). 
RetII bounds obtained in this work significantly improve limits derived in R14. 
This is despite the maps used in both works for the constraints (i.e., the $f_{15}$ tapered maps) are confusion limited and have similar synthesized beams, thus their noise is approximately equal, namely $\sim 150\,\mu$Jy.
On the other hand, what matters for the bounds on diffuse emission is the noise of the maps after subtraction of small-scale sources.

The rms noise of the $r_{-1}$ map of RetII is a factor of 3-4 lower than the maps in R14, due to the longer observational time.
This means that for RetII we have a deeper and improved characterization of background sources, which, in turn, implies their subtraction is more effective than in R14.
The noise of the resulting maps is therefore lower (by a factor $\mathcal{O}(1)$, depending on the R14 targets).
It could be further lowered by, on one hand, increasing the observational time detecting new soures, and, on the other, introducing intermediate baselines to better image bright sources. 

The lesson we learn from the comparison in Fig.~\ref{fig:bounds} is that for the near future campaigns that will be conducted with the SKA and its precursors, a special effort should be invested in performing a very good characterization of background sources and in studying effective methods to subtract them to clean low resolution maps.

In the right panel of Fig.~\ref{fig:bounds}, we report the bounds on decaying DM, for the same choice of final states as in the left panel.
Again, the level of bounds significantly improves with respect to what found in R14.

As for the case of the phenomenological model introduced in the previous Section, none of the DM models, neither in the decaying nor in the annihilating scenario, provides a $\chi^2$ value in Eq.~\ref{eq:like} smaller than for the case of a spatially constant flux density, which could have suggested an evidence for DM.

\section{Conclusions}
\label{sec:concl}

We performed a deep radio search for synchrotron emission induced by annihilation or decay of WIMPs in the Reticulum II dSph galaxy.
Observations were conducted with the ATCA telescope at 16 cm wavelength, with an rms sensitivity of 0.01 mJy/beam, and complemented on large angular scales with KAT-7 observations.
With the analysis presented in this paper, we reach the following conclusions:

\begin{itemize}
\item 240 background radio sources have been detected, with the vast majority being previously unknown. The properties of these sources are reported in the catalog described in Section~\ref{sec:cat}. 

\item Two of the detected sources are compelling BL Lac candidates and can be the radio counterpart of the possible evidence of \g-ray emission from RetII reported by Ref.~\cite{Geringer-Sameth:2015lua} using Fermi-LAT data.

\item We find no evidence for a diffuse emission from RetII.

\item Given the above point, we derive bounds on the WIMP properties. 
For a benchmark scenario, where we assume a $\mu$G magnetic field, a Milky-Way like spatial diffusion of electrons and positrons and an Einasto profile for the DM distribution, the bounds on the annihilation rate are around (below) the ``thermal'' value for hadronic (leptonic) channels and $M_\chi\lesssim 100$ GeV. 

\item Comparing with Ref.~\cite{Regis:2014tga}, we show that an improvement in the sensitivity for detection and characterization of background sources (that, for DM searches, are a noise to be subtracted) corresponds to a consequent improvement in the DM bounds.
\end{itemize}

The latter point is particularly relevant, suggesting a promising near future for radio searches of particle DM in light of the capability that will be reached by the SKA and its precursors~\cite{Colafrancesco:2015ola}.

\section*{Acknowledgements}
We warmly thank Vincent Bonnivard and David Maurin for providing us with the MCMC chains of their analysis on the Reticulum II DM profile~\cite{Bonnivard:2015tta}, and Francesco Massaro for valuable suggestions on Section~\ref{sec:excess}. 

S.C. acknowledges support by the South African Research Chairs Initiative of the Department of Science and Technology and National Research Foundation and by the Square Kilometre Array (SKA).
M.R. acknowledges support by the {\sl Excellent Young PI Grant: The Particle Dark-matter Quest in the Extragalactic Sky} funded by the University of Torino and Compagnia di San Paolo, and by the research grant {\sl TAsP (Theoretical Astroparticle Physics)} funded by the Istituto Nazionale di Fisica Nucleare (INFN).

The Australia Telescope Compact Array is part of the Australia Telescope National Facility which is funded by the Commonwealth of Australia for operation as a National Facility managed by CSIRO.

\end{document}